\documentclass{aastex61}
\usepackage{graphicx}
\usepackage{epstopdf}
\usepackage{float}
\begin{document}
\title{Magnetic Field Modeling of hot channels in Four Flare/CME Events}
\author{Tie Liu}
\affil{Key Laboratory for Dark Matter and Space Science, Purple Mountain Observatory, CAS, Nanjing 210008, China}
\affil{School of Astronomy and Space Science, University of Science and Technology of China, Hefei, Anhui 230026, China}
\author{Yingna Su}
\affiliation{Key Laboratory for Dark Matter and Space Science, Purple Mountain Observatory, CAS, Nanjing 210008, China}
\affiliation{School of Astronomy and Space Science, University of Science and Technology of China, Hefei, Anhui 230026, China}
\author{Xin Cheng}
\affiliation{School of Astronomy and Space Science, Nanjing University, Nanjing, Jiangsu, 210093,  China}
\affiliation{Key Laboratory for Modern Astronomy and Astrophysics (Nanjing University), Ministry of
	Education, Nanjing 210023, China}
\author{Adriaan van Ballegooijen}
\affiliation{5001 Riverwood Avenue, Sarasota, FL 34231, USA}
\author{Haisheng Ji}
\affiliation{Key Laboratory for Dark Matter and Space Science, Purple Mountain Observatory, CAS, Nanjing 210008, China}
\affiliation{School of Astronomy and Space Science, University of Science and Technology of China, Hefei, Anhui 230026, China}
\email{ynsu@pmo.ac.cn}
\begin{abstract}

We investigate the formation and magnetic topology of four flare/CME events with filament-sigmoid systems, in which the sigmoidal hot channels are located above the filaments, and they appear in pairs prior to eruption. The formation of hot channels usually takes several to dozens of hours during which two J-shape sheared arcades gradually evolve into sigmoidal hot channels, then they keep stable for tens of minutes or hours and erupt. While the low-lying filaments show no significant change. We construct a series of magnetic field models and find that the best-fit preflare models contain magnetic flux ropes with hyperbolic flux tubes (HFTs). The field lines above the HFT correspond to the high-lying hot channel, while those below the HFT surround the underlying filaments. In particular, the continuous and long field lines representing the flux rope located above the HFT match the observed hot channels well in three events. While for SOL2014-04-18 event, the flux bundle that mimics the observed hot channel is located above the flux rope. The flux rope axis lies in a height range of 19.8 Mm and 46 Mm above the photosphere for the four events, among which the flux rope axis in SOL2012-07-12 event has a maximum height,  which probably explains why it is often considered as a double-decker structure. Our modeling suggests that the high-lying hot channel may be formed by magnetic reconnections between sheared field lines occurring above the filament prior to eruption.
 
\end{abstract}
\section{Introduction} \label{sec:intro}
Filament eruptions, CMEs, and solar flares are different manifestations of a single physical process and the most spectacular explosive phenomena in the solar system. They eject a large quantity of plasma and magnetic flux as well as relativistic particles with intense brightness enhancement \citep{2004JGRA..109.7105Y}. According to \cite{1985SoPh..102..131T}, the most commonly seen flare morphology is two flare ribbons that usually appear in pairs and separate with time. Flare loops are located above the flare ribbons, with new magnetic field lines reconnecting at higher and higher altitudes interpreting the ribbons separation and flare loops expansion, according to the classical “CSHKP” model \citep{1992LNP...399....1S}. If the high-speed magnetized plasma and energetic particles arrive at the Earth, they will interact with the magnetosphere and ionosphere and may seriously affect the safety of satellites and astronauts in the outer space, as well as leading to damage of communications and power transport.  

Theoretical solar physicists thought that magnetic free energy stored in the corona plays a major role in driving the explosion. A magnetic flux rope (MFR) is proposed to be the fundamental structure in the flare/CME dynamical process (e.g., \citealt{1995ApJ...451L..83S, 1999A&A...351..707T, 2011LRSP....8....1C, 2013SoPh..284..179V}). Photospheric magnetic cancellations \citep{1989ApJ...343..971V}, emergence of a helical flux rope \citep{2009ApJ...697..913O} and the tether-cutting reconnection \citep{1984SoPh...94..341S} are main mechanisms for the buildup as well as the initially rising of MFR structures in the solar corona. A number of studies (e.g., \citealt{2011ApJ...738..161I, 2012ApJ...745L...4L, 2013ApJ...778L..36L, 2014ApJ...797...80V}) have presented examples on the formation of MFRs.

Filament eruptions are categorized into full eruptions, partial eruptions, and failed eruptions \citep{2007SoPh..245..287G}. “Partial eruptions” are more complicated to define observationally. The first type of partial eruption occurs when the entire magnetic structure erupts, with the eruption containing either some or none of its supported pre-eruptive filament mass. The second type of partial eruption occurs when the magnetic structure itself partially escapes with either some or none of the filament mass. \citet{2006ApJ...637L..65G} presented a three-dimensional (3D) numerical magnetohydrodynamic (MHD) simulation of a CME and find that the loss of equilibrium of a twisted flux rope results in the splitting of the rope in two, with one rope successfully being expelled and the other remaining behind. The critical factors that lead to the partially expelled rope are its three-dimensionality and its possession of dipped field lines grazing the photosphere (i.e., a bald patch). It is possible that the “degree of emergence” of a preeruption flux rope, that is, whether it possesses a bald patch (BP) or whether it is high enough in the corona to possess an X-line, determines whether the rope is expelled totally or partially. If this is indeed so, then partially erupting filaments should be more likely to possess a BP (or BPs) than totally erupting filaments.

In the past, there are mainly two groups of models for the pre-eruption magnetic configuration. One group suggests that a MFR formed above the polarity inversion line (PIL) and may keep stable for several hours and then erupt for some reason (\citealt{1991ApJ...373..294F, 1997SoPh..175..719W, 2000ApJ...539..964K}). In the other group, a MFR does not exist prior to eruption, but is formed through reconnection between two sheared arcades during the eruption (\citealt{1994ApJ...430..898M, 2003JGRA..108.1162M}). Therefore, a MFR is present during the eruption in both cases. Observations of interplanetary magnetic clouds have confirmed existence of the flux rope \citep{1991PCS....21....1B}.  Lately, the observations of a double-decker filament by \citet{2012ApJ...756...59L}  further complicate the pre-eruption magnetic configuration. They find that the active-region dextral filament is composed of two branches separated in height by about 13 Mm. A transient hard X-ray sigmoid appears between the two original filament branches during the impulsive phase of the flare. They thus suggest two types of force-free magnetic configurations that are compatible with the data, a double flux rope equilibrium and a single flux rope situated above a loop arcade.

Owing to most models of flare/CME events containing a MFR,  various observations and simulations have been carried out to find evidence of the MFRs as well as understand their formation, structure, eruption and the associated phenomena such as coronal waves and dimming. Soft X-ray (SXR) and extreme ultraviolet (EUV) observations show that forward or reversed sigmoidal structures are proxies of flux ropes in the corona (\citealt{2002ApJ...574.1021G, 2014SoPh..289.3297S}) and often considered as progenitors of CMEs. Filaments (e.g., \citealt{2010SSRv..151..333M}) and filament channels (e.g., \citealt{1985spit.conf..710G}) can also serve as indicators of MFRs. Other evidences include dark cavities at the solar limb, the descending motion of filament materials along a helical trajectory (e.g., \citealt{2014ApJ...784L..36Y, 2015A&A...580A...2Z}), spinning motions (\citealt{1995ApJ...443..818L}; \citealt{2012ApJ...752L..22L}) and  ``lagomorphic'' structure of linear polarization in cavities \citep{2013ApJ...770L..28B}. EUV observations show that ``coronal waves" and ``dimming" (\citealt{2004A&A...427..705Z, 2017A&A...598A...3Z}) are closely linked to the origins of CMEs. Upon eruptions, the magnetic loops rooted in the dimming regions (as suggested to be the locations of flux rope footpoints), which are open to the solar wind, then plasma expands and escapes along the open field lines to make the regions become dark \citep{2000GeoRL..27.1431T}. Various case studies have been carried out to analyze the formation and eruption of MFRs (e.g., \citealt{2000ApJ...532..628S, 2013AdSpR..51.1967S, 2014ApJ...797...80V}).

Lately, \cite{2012NatCo...3E.747Z} and \cite{2013ApJ...763...43C} reported that the MFR appears as a coherent hot channel structure when seen off the solar limb before the eruption (also see  \citealt{2013ApJ...778..142T, 2015ApJ...812...50J}).  \cite{Cheng2016} summarized the characteristics of the hot channel: 1) It is often observed as an EUV sigmoidal structure in the high temperature passbands of the Atmospheric Imaging Assembly. 2) In the other low temperature passbands, it appears as a dark cavity. 3) It is located above the main PIL but it’s axis has an significant deviation from the PIL. 4) One footpoint of the hot channel originates in the penumbra or penumbra edge with stronger magnetic field, while the other footpoint lies in the moss region with weaker magnetic field. Below the hot channel, a filament channel is often observed, \cite{2014ApJ...789L..35C} and \cite{2014ApJ...794..149C} concluded that the hot channel is most likely the MFR system and the prominence corresponds to cool materials in the bottom of the helical field lines. The hot channel may separate from the associated prominence and drive a CME. Furthermore, \cite{2014ApJ...780...28C} carried out a case study and identified that the hot channel can evolve smoothly from the inner into the outer corona retaining its coherence, its morphology coincides with the CME cavity in the white-light images. \cite{2015ApJ...808..117N} performed a statistical study and found that almost half of major eruptive flares contain a hot channel-like structure. All of these studies support the idea that EUV hot channels or blobs correspond to MFRs.

\cite{2017arXiv170605769G} reviewed the 3D magnetic field models which include theoretical force-free field models, numerical nonlinear force-free field models and MHD models. Theoretical force-free field models contain potential field in the cartesian coordinate system (e.g., \citealt{1964NASSP..50..107S}) and the spherical coordinate system (e.g., \citealt{1969SoPh....9..131A}) as well as nonlinear force-free field (NLFFF) models (e.g., \citealt{1990ApJ...352..343L}). Numerical NLFFF models are constructed from boundary conditions and proper initial conditions. There are various numerical algorithms to compute NLFFF, such as the Grad-Rubin, vertical integration, MHD relaxation, optimization, and boundary integral equation methods \citep{2012LRSP....9....5W}. In order to study the dynamics of a magnetic flux rope, we need MHD numerical simulations which can be divided into zero-$\beta$, isothermal, ideal, resistive, and full MHD models with the order of increasing physical details included and can also be divided into purely theoretical simulations (\citealt{2000ApJ...529L..49A, 2012ApJ...744...66Z}) and data-driven/data-constrained simulations (\citealt{2013ApJ...771L..30J, 2016NatCo...711522J, 2014Natur.514..465A,  2015ApJ...803...73I}) according to the adopted initial and boundary conditions. 

The Coronal Modeling System (CMS; \citealt{2004ApJ...612..519V} ) used to model non-potential magnetic fields in the corona has been proven to be effective in studies of the magnetic field prior to the eruption, such as active regions (\citealt{2008ApJ...672.1209B}; \citealt{2009ApJ...691..105S,2009ApJ...704..341S,2011ApJ...734...53S}; \citealt{2018ApJ...855...77S}), coronal X-ray sigmoids (\citealt{2009ApJ...703.1766S, 2015ApJ...810...96S}), and polar crown prominences \citep{2012ApJ...757..168S, 2015ApJ...807..144S}. Different from usually used extrapolation methods which extrapolate the observed photospheric vector field into the corona (e.g., \citealt{2010ApJ...715.1566C, 2010ApJ...714..343G, 2016ApJ...823...62Z}), CMS inserts a MFR into the potential-field of an active region then applies magneto-frictional relaxation to evolve the field into NLFFF or unstable states (\citealt{1986ApJ...309..383Y}; \citealt{2000ApJ...539..983V}), so it is also referred as ``flux rope insertion method". The shape of the inserted MFRs is constrained by observed filaments, filament channels or PIL. The axial flux (along the axis) as well as  poloidal flux (ringing around the axis) of the MFR can be adjusted as initial conditions. ``Flux rope insertion method" can achieve large coronal volume in and around the target region and high spatial resolution in the lower corona owing to variable grid spacing. 

In this work we study four flare/CME events with filament-sigmoid systems which contain a hot channel located above a filament near the disk center. The eruption of the hot channel leads to a CME while the low-lying filament remains fully or partly. Observational studies of these four events have been presented by \cite{Cheng2016}. Other detailed information about the four events can be obtained from the literature \citep[e.g.,][]{2014ApJ...789L..35C, 2014ApJ...784..144D, 2015ApJ...810...96S, 2017ApJ...845...26J, 2015ApJ...812...50J, 2016ApJ...823...41D, 2016ApJ...823...62Z, 2017A&A...598A...3Z}.  Observations and  extrapolations suggest that EUV hot channels and filament channels are most promising evidence of MFRs. Case studies using NLFFF extrapolations like  \cite{2016ApJ...823...62Z} have shown good correspondence of the hot channel and the MFR. Using ``flux rope insertion method", we revisit these four events in order to understand the detailed magnetic topology as well as the formation mechanism of these systems. In Section \ref{ sec:INSTRUMENTS}, we introduce data and instrumentation. In Section \ref{sec: Observations}, we briefly review the observations. Results of magnetic modeling are presented in Section \ref{ sec: modeling}. Section \ref{sec:SUMMARY AND DISCUSSION} presents summary and discussions.

\section{ INSTRUMENTS} \label{ sec:INSTRUMENTS}
	\begin{deluxetable*}{lcc}
		\tablecaption{AIA channels. \label{tab:aia}}
		\tablecolumns{3}
		\tablenum{1}
		\tablewidth{0pt}
		\tablehead{
			\colhead{Channel Name} & \colhead{Primary Ion(s)} & \colhead{Char. T (MK)}
				}	
		\startdata
		131 {\AA} & Fe \uppercase\expandafter{\romannumeral8}, \uppercase\expandafter{\romannumeral21}  & $\backsim$0.4, 10\\
		\hline
		94 {\AA} & Fe \uppercase\expandafter{\romannumeral18} & $\backsim$6.3\\
		\hline
		335 {\AA} & Fe \uppercase\expandafter{\romannumeral16} & $\backsim$2.5\\
		\hline	
		211 {\AA} & Fe \uppercase\expandafter{\romannumeral14} & $\backsim$2\\
		\hline	
		193 {\AA} & Fe \uppercase\expandafter{\romannumeral12} & $\backsim$1.6\\
		\hline	
		171 {\AA} & Fe \uppercase\expandafter{\romannumeral9} & $\backsim$0.6\\
		\hline	
		304 {\AA} & He \uppercase\expandafter{\romannumeral2} & $\backsim$0.05\\
		\hline							
		\enddata	
	\end{deluxetable*}	
Observational data are taken with Solar Dynamics Observatory/Atmospheric Imaging Assembly (SDO/AIA; \citealt{2012SoPh..275...17L}) and SDO/Helioseismic Magnetic Imager (SDO/HMI; \citealt{2012SoPh..275..229S}). AIA has a large field of view ($>$ 1.3 solar diameters) and ten visible channels (7 extreme ultraviolet (EUV) and 3 ultraviolet (UV))  with 12 s or 24 s cadence and $0\arcsec.6$ pixel size. The primary ions and emission temperature for 7 EUV passpands are listed in Table \ref{tab:aia}. HMI observes the full solar disk in 6173 {\AA}, it provides light-of-sight magnetograms and vector magnetograms with time cadences of 45 s/720 s and pixel size of $0\arcsec.5$.

\section{ Observations: Formation and structure of the filament-sigmoid systems } \label{sec: Observations}

\begin{figure*}
	\centering \includegraphics[width=7in]{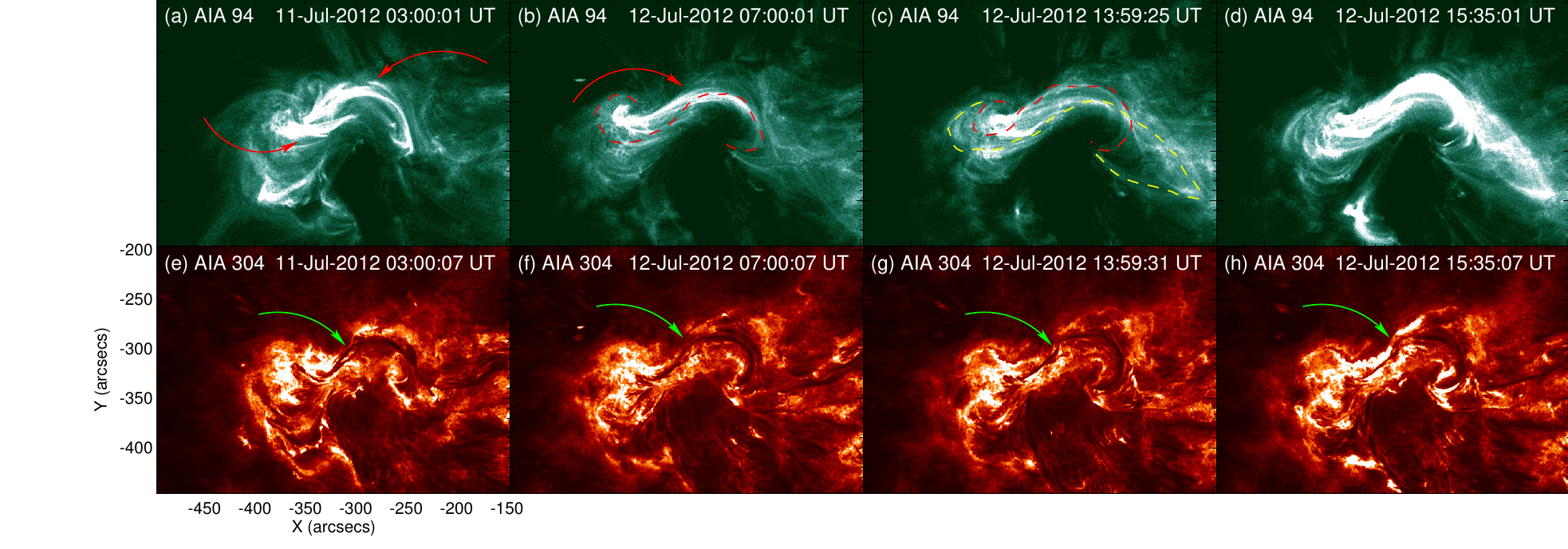}
	\caption{(a)-(d) SDO/AIA 94 {\AA} and (e)-(h) 304 {\AA} images  show the formation of the SOL2012-07-12T hot channel. Red arrows mark the initial sheared arcades and the main reconnection area, green arrows mark the filament channel. The low-lying part of the double-decker is traced with a red dashed line and the high-lying part is traced with a yellow dashed line.} 
	\label{fig: evo1} 
	
\end{figure*}
\begin{figure*}
	\centering \includegraphics[width=7in]{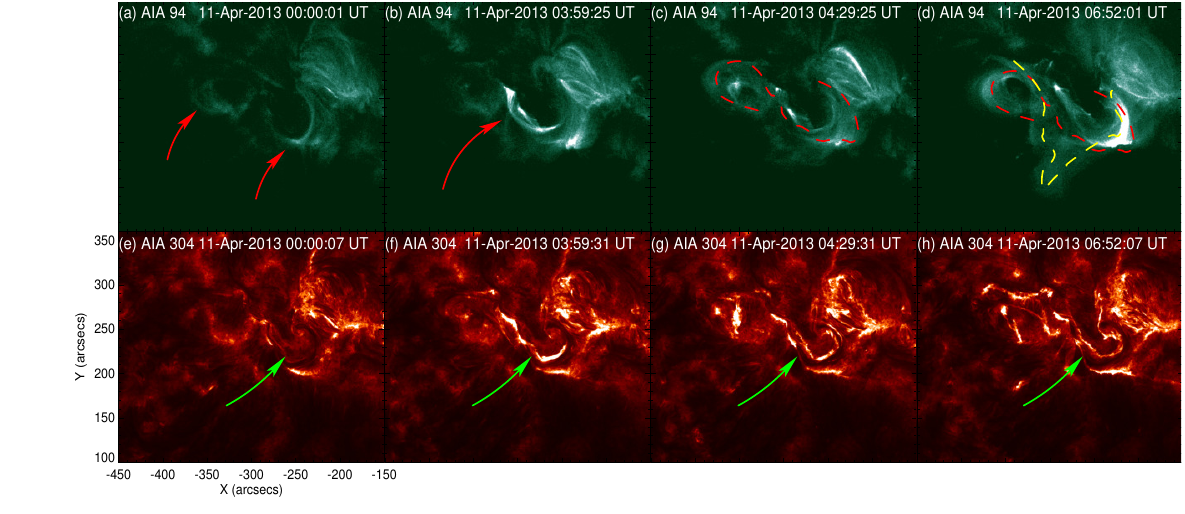}
	\caption{Red dashed line traces the hot channel and yellow dashed line traces the overlying flux boundle, others are the same as Figure \ref{fig: evo1} but for the SOL2013-04-11T event.} 	
	\label{fig: evo2} 
\end{figure*}
\begin{figure*}
	\centering \includegraphics[width=7in]{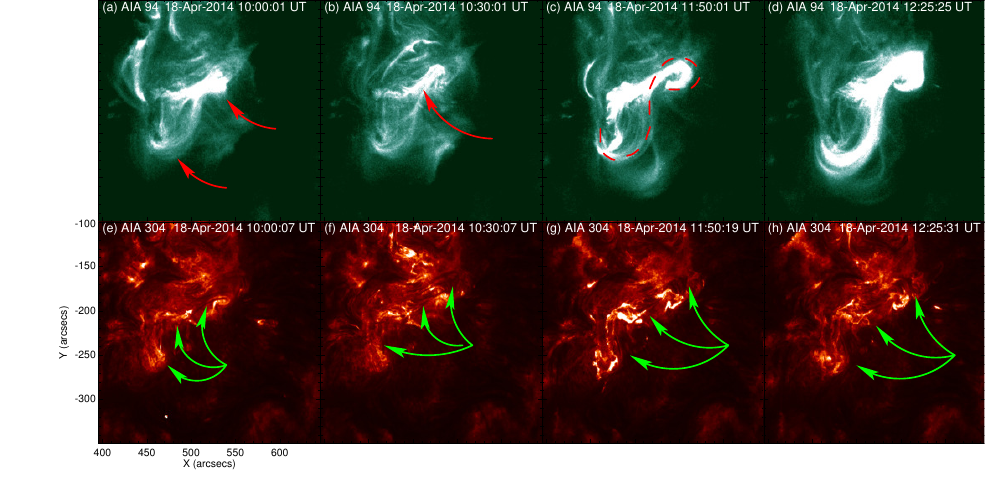}
	\caption{Three green arrows mark the filament channel others are the same as Figure \ref{fig: evo1} but for the SOL2014-04-18T event.} 	
	\label{fig: evo3} 
\end{figure*}
\begin{figure*}
	\centering \includegraphics[width=7in]{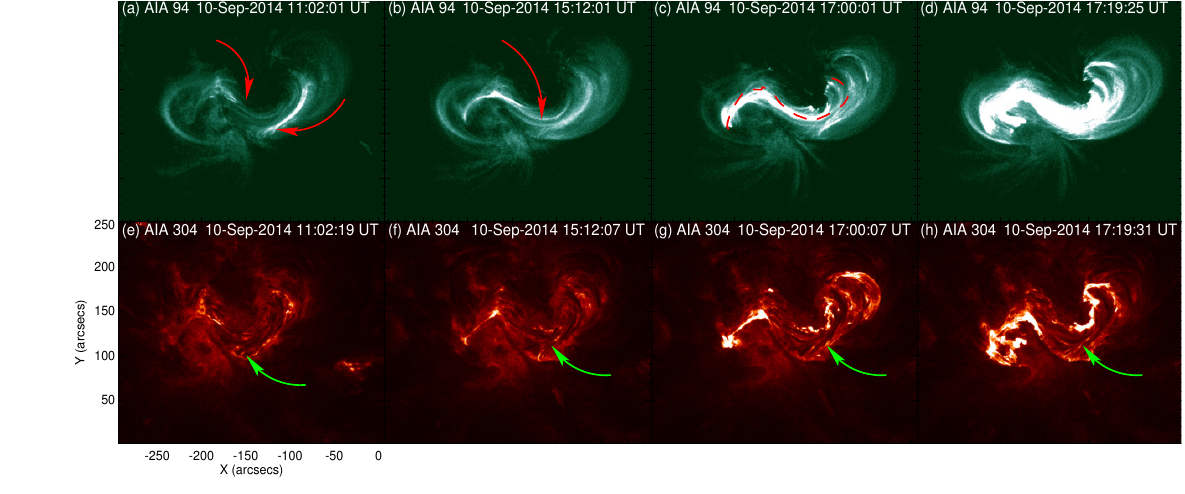}
	\caption{Same as Figure \ref{fig: evo1} but for the SOL2014-09-10T event.} 	
	\label{fig: evo4} 
\end{figure*}

Observational studies on the formation of each of the four filament-sigmoid systems can be found in the literature \citep[e.g.,][]{Cheng2016}.  In this section, we briefly review the observations. Figure \ref{fig: evo1} presents the formation of the hot channel in the first event (SOL2012-07-12T) as shown in the AIA 94 {\AA} and 304 {\AA} images.  AIA 94 {\AA} images show that the hot core structure initially appears as two J-shape loops marked with red arrows in panel a, then evolves into a continuous S-shape hot channel traced with a red dashed line in panel b. At 07:00 UT the middle part lights up (marked with a red arrow in panel b), which suggests magnetic reconnecton may take place. \cite{2014ApJ...789...93C} finds that the entire sigmoid furthermore evolves into a double-decker MFR structure, the high-lying one (traced with a yellow dashed line in panel c) and the low-lying one (traced with a red dashed line in panel c) manifests as the hot channel and the filament, respectively. The system keeps stable (panel c) for a few hours and then its high-lying part becomes unstable (panel d) and erupts leading to a CME, while the low-lying part does not erupt. In the meanwhile, the AIA 304 {\AA} images show that the low-lying filament (marked with a green arrow) roughly keeps stable. 

For the second event (SOL2013-04-11T: Figure \ref{fig: evo2}), we obtain similar impression. At 03:59 UT there are brightenings at the middle part (marked with a red arrow in panel b), then the hot channel (traced with a red dashed line in panel c) is formed due to the merging of two J-shape sheared arcades marked by red arrows in panel a. Unlike the first event, this region seems to contain two hot channels. When it becomes unstable (panal d), the high-lying hot channel traced with yellow dashed line starts to kink and then erupts, followed by the eruption of the low-lying sigmoidal hot channel traced with red dashed line as suggested by  \citet{2017ApJ...834...42J, 2014ApJ...797...80V}. From the AIA 304 {\AA} images we find that the left part of the filament marked with a green arrow is not as evident as that in the right part associated with brightenings in panel d. The eruption only leads to the disappearance of the hot channels while the low-lying filament appears to remain there at least partly. 

The third event (SOL2014-04-18T) is a good example of hot channel formation due to tether-cutting reconnection. Analyzing IRIS spectral data and images of SDO, \cite{2015ApJ...804...82C} concludes that the formation of the continuously sigmoidal loops in the hot channel is mainly via tether-cutting reconnection in the chromosphere. \cite{2015ApJ...812...50J} also agrees that tether-cutting reconnection plays a major role. Figure \ref{fig: evo3} presents the formation of the hot channel, just like the first event, two sheared arcades (marked with red arrows in panel a) gradually evolve into long S-shape threads traced with a red dashed line in panel c. When the structure becomes unstable (panel d), the eastern part of the brightening moves towards the south with the rising of the high-lying hot channel. No clear filaments can be identified in the corresponding 304 {\AA} images except several dark filamentary threads marked with green arrows. 

For the  SOL2014-09-10T event (Figure \ref{fig: evo4}), the left sheared loops together with the right sheared loops (red arrows in panel a) make up of the whole sigmoidal structure (red dashed line in panel c).  In the early phase of the eruption (panel d), the hot channel is already unstable and long S-shape loops are formed at the sigmoid center. Panels a and b show that brightenings take place at the right and the left parts of the hot channel respectively, which suggests that the reconnection occurs not only between the two J-shape loops but also in the surroundings. Using IRIS spectral lines, \cite{2015ApJ...804...82C} find that the reconnection occurs at a lower height. The low-lying filament consists of two sections, and the western section undergoes failed eruption, while most parts of the filament remain unchanged \citep{2016ApJ...823...41D}. In the end, the high-lying hot channel erupts, while the low-lying filament stays behind. 

The target four filament-sigmoid systems appear as forward (SOL2012-07-12T and SOL2014-04-18T events) and reversed (SOL2013-04-11T and SOL2014-09-10T events) sigmoids. Observations suggest that the low-lying filament is formed firstly.  Prior to the eruption two groups of sheared arcades gradually evolve into a long sigmoidal hot channel located above the preexisiting filament during tens of (or several) hours in the AR. Once the hot channel is formed, it usually remains stable for a while then erupts suddenly, while the low-lying filaments remain fully or partially stable. During this time interval the associated filaments in three events show no clear changes.

\section{Modeling} \label{ sec: modeling}

	\begin{deluxetable*}{lccccccc}
		\tablecaption{Caracteristics of the Four Events. \label{tab:cha}}
		\tablecolumns{8}
		\tablenum{2}
		\tablewidth{0pt}
		\tablehead{
			\colhead{Case} & \colhead{Position} & \colhead{Shape}& \colhead{Flare}& \colhead{L-path}& \colhead{H-x} & H-FR & \colhead{H-null}
		}	
		\startdata
		2012-07-12 & S13W03 & S-shape & X1.4 & $\backsim158.9 Mm$ & $\backsim16.9 Mm$ &  $\backsim46.0 Mm$&$\backsim50.5 Mm$\\
		\hline
		2013-04-11 & N07E13 & I-S-shape & M6.5 & $\backsim232.4 Mm$ & $\backsim8.4 Mm$ &  $\backsim34.2 Mm$ &None\\
		\hline
		2014-04-18 & S20W34 & S-shape & M7.3 & $\backsim293.3 Mm$ & $\backsim9.8 Mm$ &  $\backsim35.7 Mm$ &$\backsim82.5 Mm$\\
		\hline	
		2014-09-10 & N11E05 & I-S-shape & X1.6 & $\backsim182.0 Mm$ & $\backsim4.9 Mm$ &  $\backsim19.8 Mm$ &None\\
		\hline	
		\enddata	
		\tablecomments{Observational characteristics include dates, positions and shapes of hot channels, and classes of flares are listed in the left four columns.  The length of the inserted flux ropes, the heights of the HFTs and the flux rope axis, as well as the overyling null points (if exist) in the best-fit model for each event are presented in the right four columns.}
	\end{deluxetable*}	

	\begin{deluxetable*}{*{6}{c}}
		\tablecaption{Parameters of Models for the Four Events. \label{tab:par}}
		\tablecolumns{6}
		\tablenum{3}
		\tablewidth{0pt}
		\tablehead{
			\colhead{Case} & \colhead{Models} & \colhead{Axial flux}& \colhead{Poloidal flux}& \colhead{Helicity}& \colhead{Free energy}\\
			\colhead{}& \colhead{}& \colhead{$10^{20} Mx$}& \colhead{$10^{10}$ $Mx/cm$}& \colhead{$10^{42}$ $Mx^2$}& \colhead{$10^{32} erg$}		
		}
		\startdata
		2012& best-fit model & 55 & $-10$& 76.39& 10.46\\
		$\mid$& stable model 1& 20& 0& 36.27& 4.05\\
		07& stable model 2& 40& 0& 67.31& 9.98\\
		$\mid$& marginally unstable model& 50& 0& 80.77& 11.95\\
		12& unstable model& 60& 0& 93.03& 13.29\\
		\hline
		2013& best-fit model & 15 & 0& $-6.98$& 1.04\\
		$\mid$& stable model 1& 5& 0& $-2.72$& 0.19\\
		04& stable model 2& 10& 0& $-4.98$& 0.66\\
		$\mid$& marginally unstable model& 15& 0& $-6.98$& 1.04\\
		11& unstable model& 20& 0& $-8.60$& 1.19\\
		\hline
		2014& best-fit model & 15 & 0& 17.19& 2.77\\
	    $\mid$& stable model 1& 5& 0& 6.33& 0.64\\
		04& stable model 2& 10& 0& 12.25& 1.68\\
		$\mid$& marginally unstable model& 15& 0& 17.19& 2.77\\
		18& unstable model& 20& 0& 21.79& 3.58\\
		\hline
		2014& best-fit model & 25 & 10& $-23.35$& 5.24\\
		$\mid$& stable model 1& 10& 0& $-11.81$& 2.22\\
		09& stable model 2& 15& 0& $-17.59$& 3.77\\
		$\mid$& marginally unstable model& 20& 0& $-23.15$& 5.23\\
		10& unstable model& 25& 0& $-28.51$& 6.57\\
		\hline
  		\enddata
		\tablecomments{Parameters for the best-fit model and four models from stable to unstable for each event. Columns 3-6 present the initial axial flux and poloidal flux of the inserted flux ropes as well as the total coronal magnetic helicity and magnetic free energy in the whole computational domain of each model.}
	\end{deluxetable*}

\begin{figure*}
	\centering \includegraphics[width=7in]{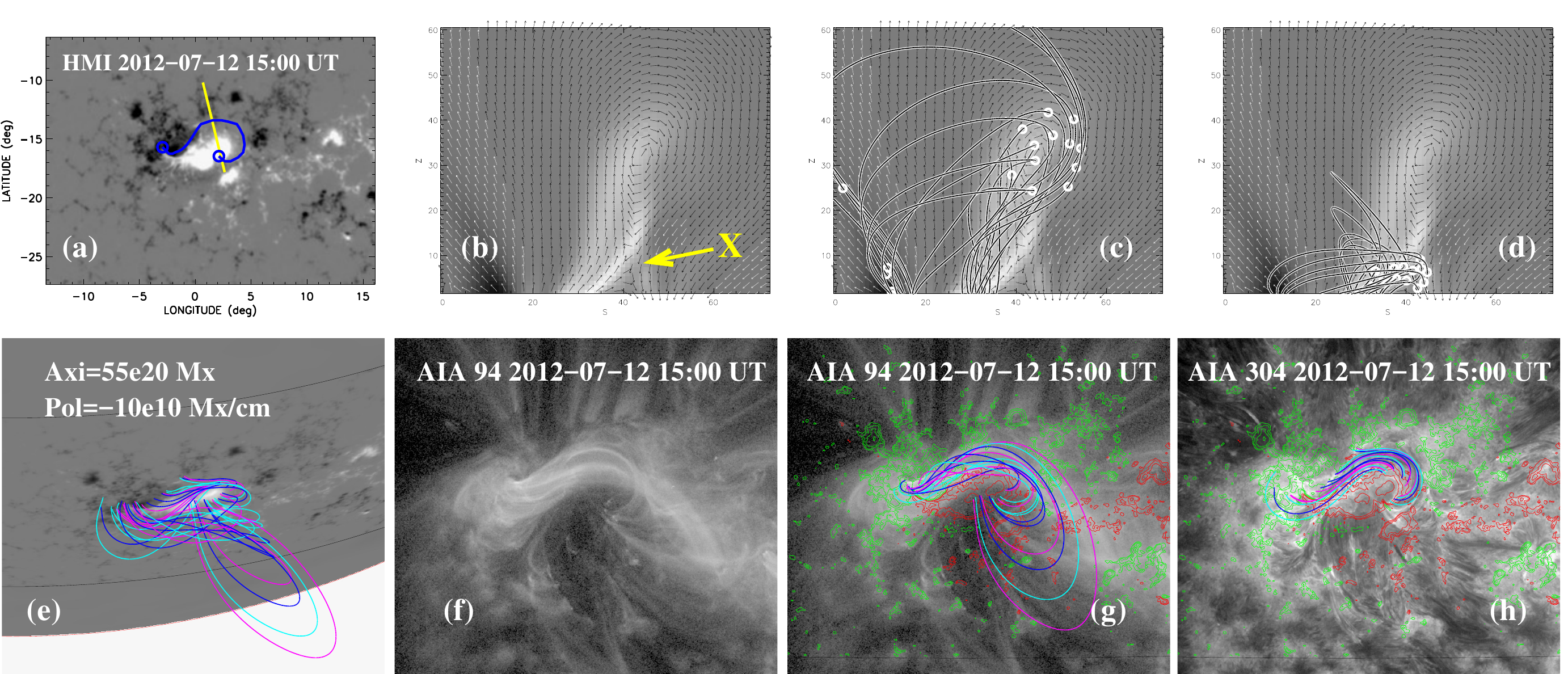}
	\caption{Pre-flare magnetic field structure at 15:00 UT on 12 July 2012. (a) The boundary condition of the best-fit model, i.e., the longitude-latitude map of the radial components of the photospheric magnetic fields in the HIRES region. The blue curve terminating with two circles represents the path along which we insert the flux rope. (b)-(d) Vertical slices of the distribution of electric currents (bright area) overlaid with magnetic vectors (black and white arrows) along the yellow line marked in panel (a). The white area represent the region with strong electric currents. In order to show the vectors well, we use black arrows on brighter background and white arrows on darker background. The yellow arrow in panel (b) marks the position of the X-point. Panels (c) and (g) show selected magnetic field lines above the X-point and magnetic field lines below the X-point are displayed in panels (d) and (h). Both selected field lines below and above the X-point are shown in panel (e) in another view. The background of panel (e) refer to the HMI LOS magnetograms, and the HIRES regionis enclosed with the black rectangle.  AIA image in 94 {\AA} and 304{\AA} at 15:00 UT is displayed in (f)-(h) as backgorund. The red and green contours representing positive and negative polarities of photospheric magnetic fields taken by HMI at 15:00 UTs, respectively. The cell size of the model  is $\backsim$1.4 Mm. } 
	\label{fig: mstr1} 
\end{figure*}

\begin{figure*}
	\centering \includegraphics[width=7in]{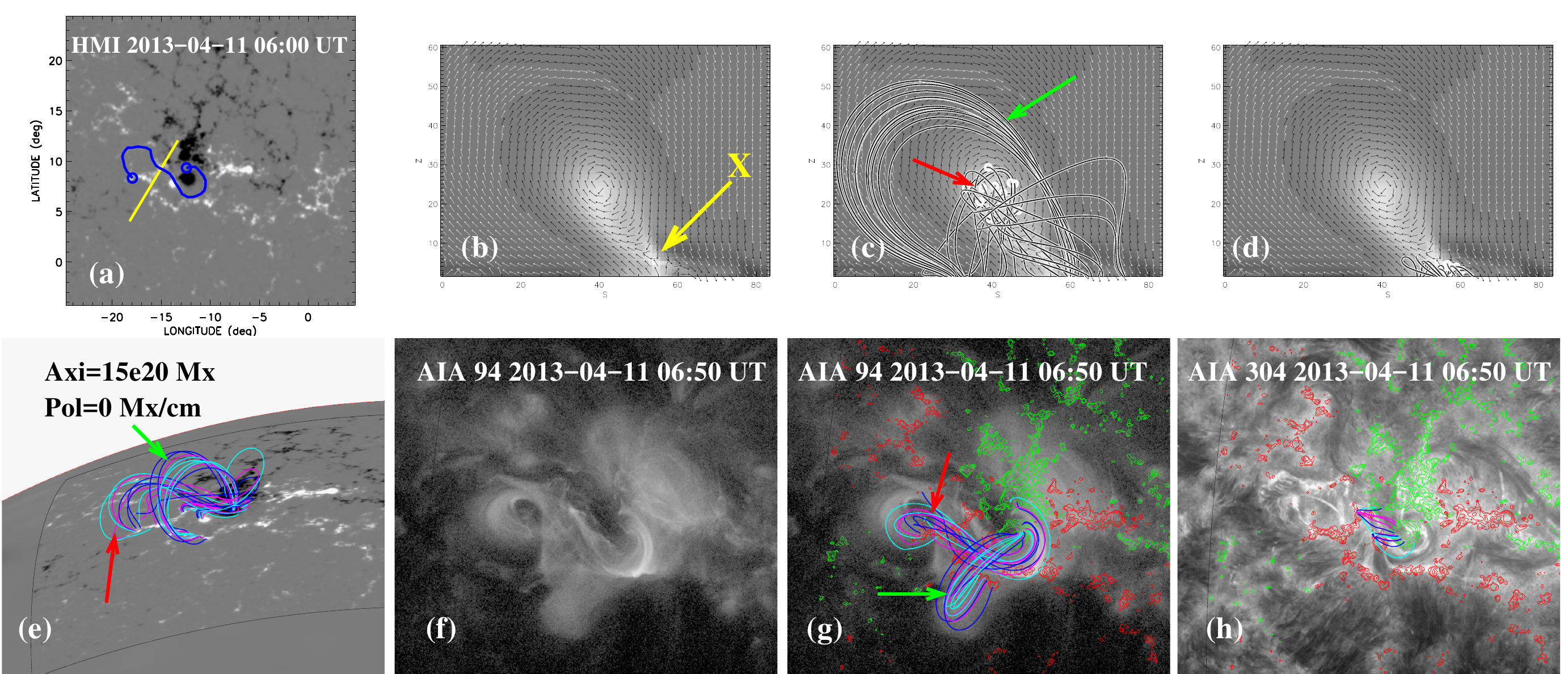}
	\caption{Same as Figure \ref{fig: mstr1}  but for the SOL2013-04-11T event at 06:49 UT. The red and green arrows refer to the field lines corresponding to the observed hot channel and the overyling hot blob, respectively.} 	
	\label{fig: mstr2} 
\end{figure*}

\begin{figure*}
	\centering \includegraphics[width=7in]{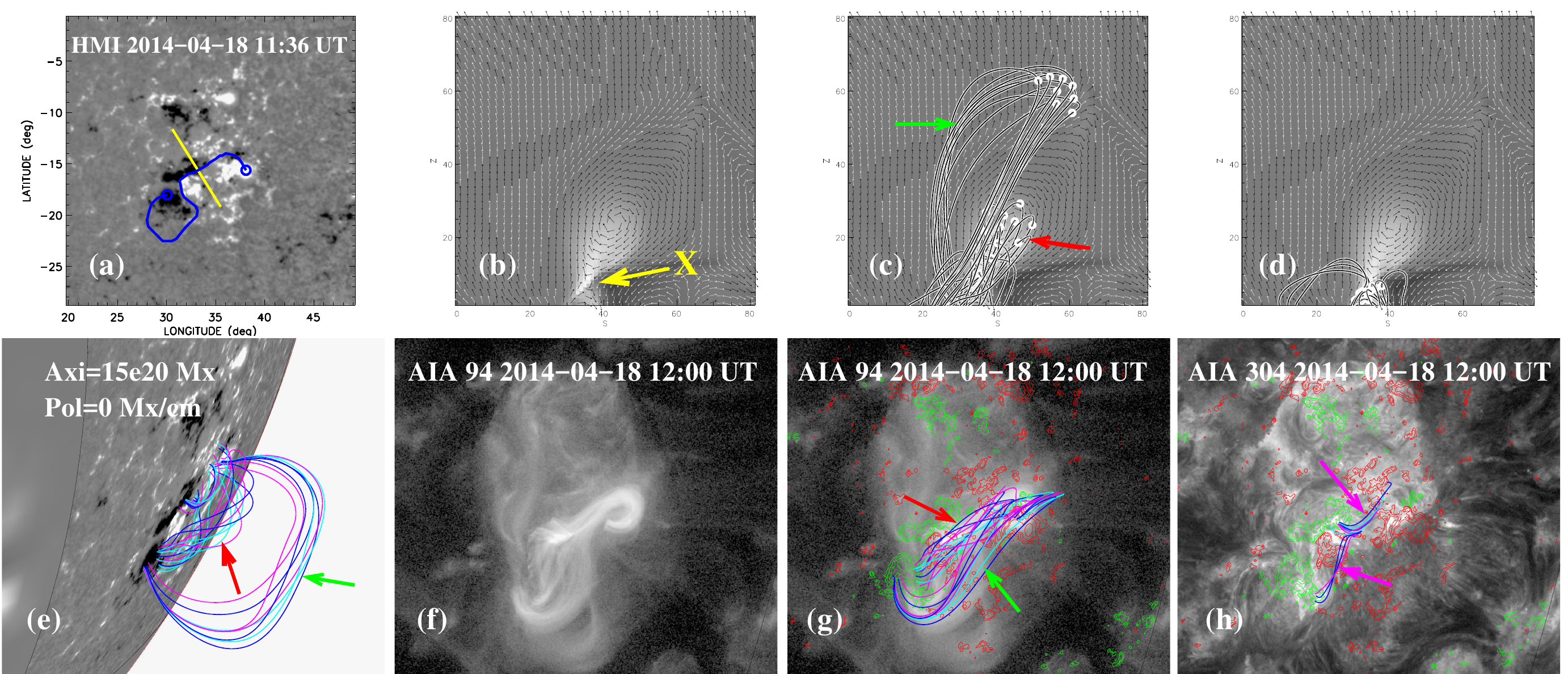}
	\caption{Same as Figure \ref{fig: mstr1} but for the SOL2014-04-18T event at 12:00 UT. The red and green arrows refer to the modeled flux rope and the overlying flux bundle that mimics the observed hot channel, respectively.} 	
	\label{fig: mstr3} 
\end{figure*}

\begin{figure*}
	\centering \includegraphics[width=7in]{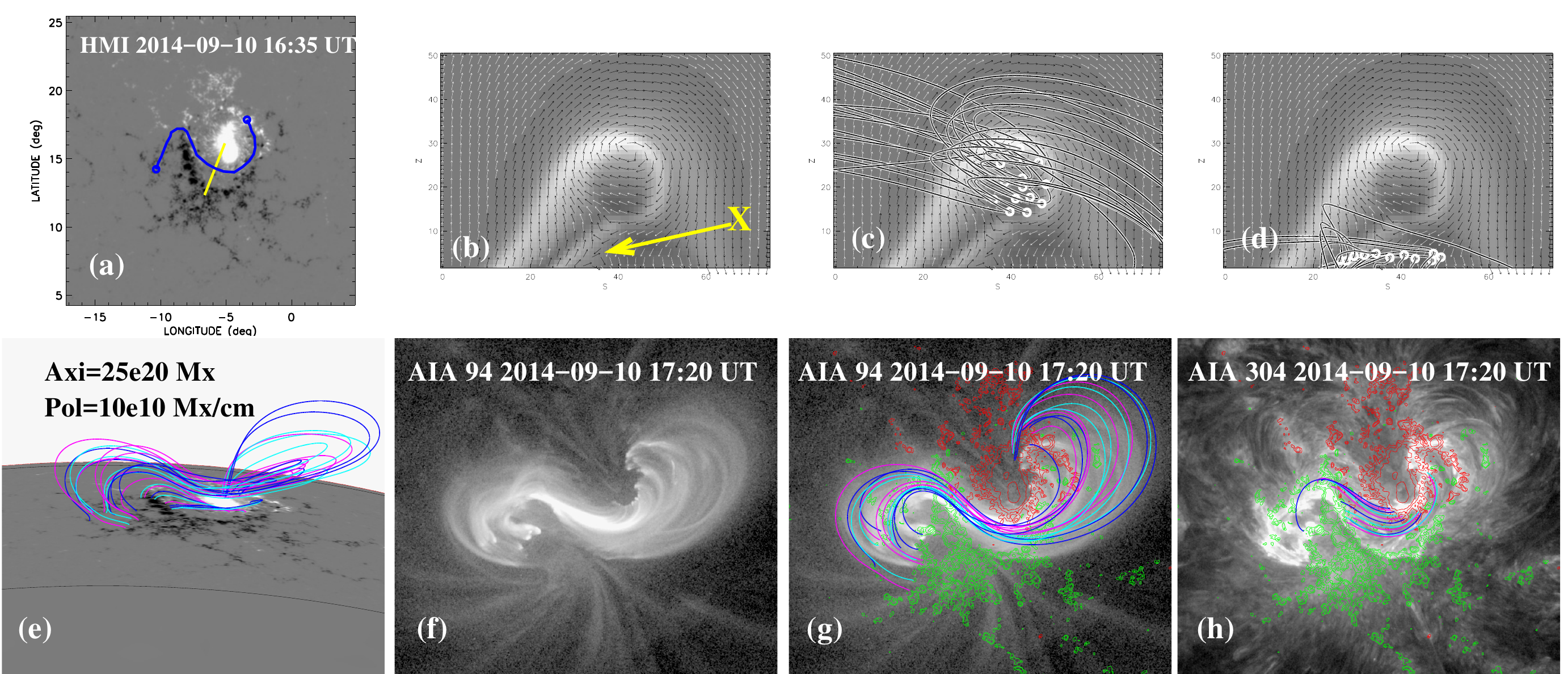}
	\caption{Same as Figure \ref{fig: mstr1}, but the cell size of this model is $\backsim$0.7 Mm for the SOL2014-09-10T event at 17:20 UT.} 	
	\label{fig: mstr4} 
\end{figure*}

\subsection{Flux Rope Insertion Method}

It is generally accepted that filament eruptions, flares and CMEs  are phenomena in the corona and the coronal magnetic field plays an important role during the eruption.  Unfortunately, we usually cannot measure the coronal magnetic field directly, although recently some progress has been made (see, e.g., \citealt{1998ApJ...500.1009J}; \citealt{2003AGUFMSH41D..05S}; \citealt{2004ApJ...613L.177L}).  These direct measurements are only available for a few individual cases and usually one has to extrapolate the coronal magnetic field from photospheric magnetic measurements. For more details, please refer to \cite{Wiegelmann_2008_JournalofGeophysicalResearch(SpacePhysics)_113_A03S02}. Different from most of extrapolation methods, we use the flux rope insertion method developed by \cite{2004ApJ...612..519V} to reconstruct NLFFF. This method only requires line-of-sight (LOS) photospheric magnetograms, while the aforementioned extrapolation methods usually require vector magnetic fields as boundary conditions. In particular, the flux rope insertion method involves inserting a magnetic flux rope into the 3D potential field then uses magneto-frictional relaxation to drive the field to force-free equilibrium. The model contains the region of target (the HIRES region e.g., an active region) with high spatial resolution (0.001 $R_\odot$ for SOL2014-09-10T event and 0.002 $R_\odot$ for the other three events at low corona) and more distant surrounding  region (GLOBAL region : global potential field ) with lower resolution ($\backsim1^\circ$). The HIRES region is reconstructed based on the HMI LOS magnetogram and the GLOBAL region is extrapolated using the HMI synoptic map. 

Four steps are required to reconstruct coronal magnetic fields  for a target  active region: 1) Extrapolating the potential field based on the observed LOS magnetogram. 2) Creating a cavity in the potential field model then inserting magnetic flux ropes  along the selected filament paths according to observations.  3) Adjusting axial flux (along the axis) and poloidal flux (circling around the axis) of the inserted magnetic flux rope (MFR) to create a grid of models. 4)  Starting magneto-frictional relaxation (\citealt{1986ApJ...309..383Y}) to evolve the models into NLFFF or unstable states, which are compared to observations to find the best-fit model. Consequently, this method produces 3D magnetic fields heavily constrained by observations.  \cite{2011ApJ...734...53S} found that increasing the axial flux or poloidal flux  of the inserted flux rope can lead to instability, hence there are critical values for both the axial flux and poloidal flux. If the axial flux or the poloidal flux is larger than the critical value the model will evolve into unstable state, otherwise it will remain stable. The marginally unstable model is located between the stable and unstable states.

NLFFF models with partial double-decker configurations (in which only part of the double-decker flux ropes shares the same PIL, e.g., one flux rope is longer than the other one) have been constructed using both extrapolations \citep{2016ApJ...826..119L,2016ApJ...818..148L,2018ApJ...857..124A} and flux rope insertion method \citep{2016NatCo...711837X, 2017ApJ...844...70L}. However, we haven't been able to produce double-decker configurations with two stable flux ropes sharing exactly the same PIL, as proposed by \citet{Liu_2012_TheAstrophysicalJournal_756_59}. We have attempted to produce such a configuration by inserting two twisted flux ropes into the model, one located above the other, but we find that this does not lead to the desired result. The reason is that during magneto-frictional evolution the flux ropes coalesce, so the relaxed model invariably contains only a single flux rope (hyper-diffusion is imposed during the relaxation process to produce a smooth magnetic field). In our models the hyperbolic flux tube (HFT) typically lies at a height of only 7 to 12 grid points, which is difficult to accomodate a stable, weakly twisted flux rope below the HFT. Vector-field extrapolation methods are also unable to reproduce double-decker configurations for our studied events in the literature, so this problem is not unique to our model. In future modeling with higher spatial resolution it may become possible to produce double-decker configurations, but for now we have to accept this limitation of our method. Therefore, in the present work we focus on reproducing the larger flux rope associated with the hot channel above the HFT, and do not attempt to produce a second flux rope in the region below the HFT.

\subsection{Structure of the Filament-Sigmoid Systems}\label{subsec:str}

In order to find the best-fit model prior to the flare, we construct a series of models (Table \ref{tab:par}) through adjusting axial flux and/or poloidal flux. Our models are constructed based on the LOS magnetograms taken tens of minutes before the flare, at this time the hot channels have been formed and are close to the unstable state. Tens of minutes later they will erupt. Detailed model parameters are listed in Tables \ref{tab:cha} and \ref{tab:par}. We find that the magnetic free energy in the best-fit model for each event is larger than $10^{32}$ erg, which is sufficient to power the observed major flares ($>$M6.0).  Model results for the four systems are presented in Figures \ref{fig: mstr1}-\ref{fig: mstr4}. Panel a in Figures \ref{fig: mstr1}-\ref{fig: mstr4}  shows longitude-latitude maps of the radial components of the photospheric magnetic fields in the HIRES region for the four events. The blue curves terminating with two circles represent the path along which we insert flux ropes. The yellow lines refer to the locations of vertical slices of electric currents displayed in panels b-d of Figures \ref{fig: mstr1}-\ref{fig: mstr4}. Selected model field lines above (panels c, g) and below (panels d, h) the X-point at a view different from the observations are shown in panel e of Figures \ref{fig: mstr1}-\ref{fig: mstr4}. AIA images in 94 {\AA} and 304 {\AA} are presented in panels f-g and panel h of Figures \ref{fig: mstr1}-\ref{fig: mstr4}, respectively. 

For all of the four events, the best-fit models contain an MFR  with HFT (X-point)  as shown in the vertical slices of electrical currents' distribution (panel b of Figures \ref{fig: mstr1}-\ref{fig: mstr4}). The height of the X-point ranges from 4.9 Mm to 16.8 Mm, and the height of the flux rope axis is located in a range between 19.8 Mm and 46 Mm as shown in Table \ref{tab:cha}. Magnetic reconnection occurs at the X-points between the two parts, the high-lying part may separate from the low-lying part and erupts, and the low-lying part of the double-decker stays behind. Observations show similar scene: the high-lying hot channel erupts and forms a CME while the low-lying filament stays in the original location. The observed hot channels correspond to the field lines above the X-point, while the low-lying filaments are surrounded by the field lines below the X-point as shown in panels c, d, g and h of Figures \ref{fig: mstr1}-\ref{fig: mstr4}. For three events (SOL2012-07-12T, SOL2013-04-11T, SOL2014-09-10T), the  long and continuous field lines representing the flux ropes located above the X-point can mimic the observed S-shaped hot channels, while the intermittent and shorter sheared field lines below the X-point appear to surround the underlying filament. In particular, there are two groups of flux bundles in the best-fit model of SOL2013-04-11T event. From Figures \ref{fig: mstr2}e-g we see that the S-shape hot channel is consistent with the S-shape flux rope above the X-point, while the first erupted overlying hot blob \citep{2014ApJ...797...80V} matches the overlying field lines above the flux rope. However, for SOL2014-04-18T event, the observed hot channel appears to be consistent with the flux bundle (green arrows) located above the flux rope near the overlying null point as shown in Figure \ref{fig: mstr3}. Typical tether-cutting reconnection configuration is shown in Figures \ref{fig: mstr3}g-h in which two J-shape sheared arcades ( pink arrows) evolve into one MFR (red arrows).  

\subsection{Topology of the Filament-Sigmoid Systems}\label{subsec:topo}

\begin{figure*}
	\centering \includegraphics[width=7in]{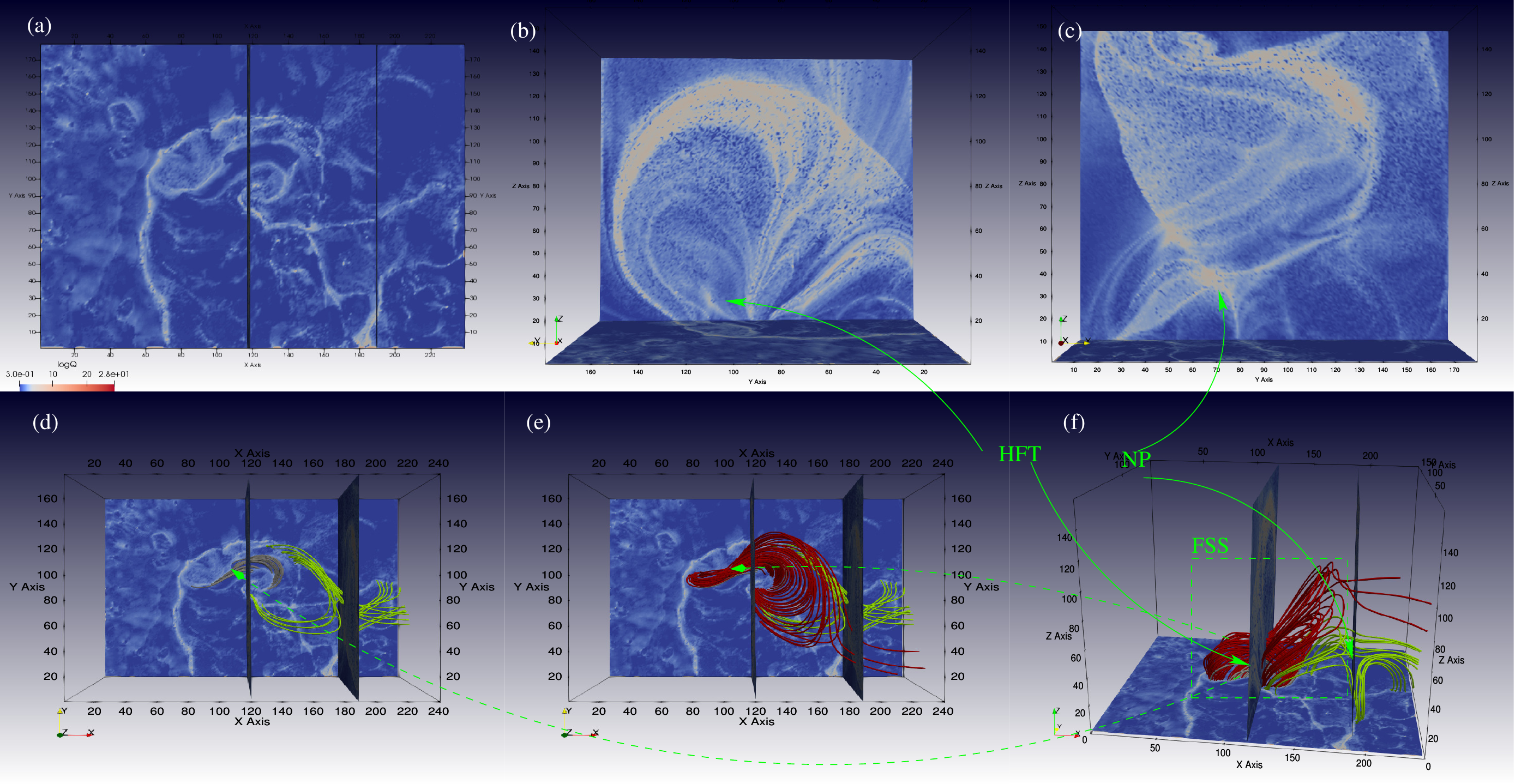}
	\caption{QSL maps overlaid with selected magnetic field lines from the best-fit model for SOL-2012-07-12 event. Panel a shows top view of the photospheric QSL map with two slits, the vertical slices along which are displayed in panels b and c. 3D combined views from the top and the side are shown in the bottom row. The HFT and null point marked by green arrows are located in panels b and c respectively. Magnetic field lines below (gray) and above the HFT (red) as well as those near the overlying null point (green) are displayed in panels d, e and f. The unit is the cell size of the model which are the same as Figure \ref{fig: mstr1}.} 	
	\label{fig: qslf1} 
\end{figure*}
\begin{figure*}
	\centering \includegraphics[width=7in]{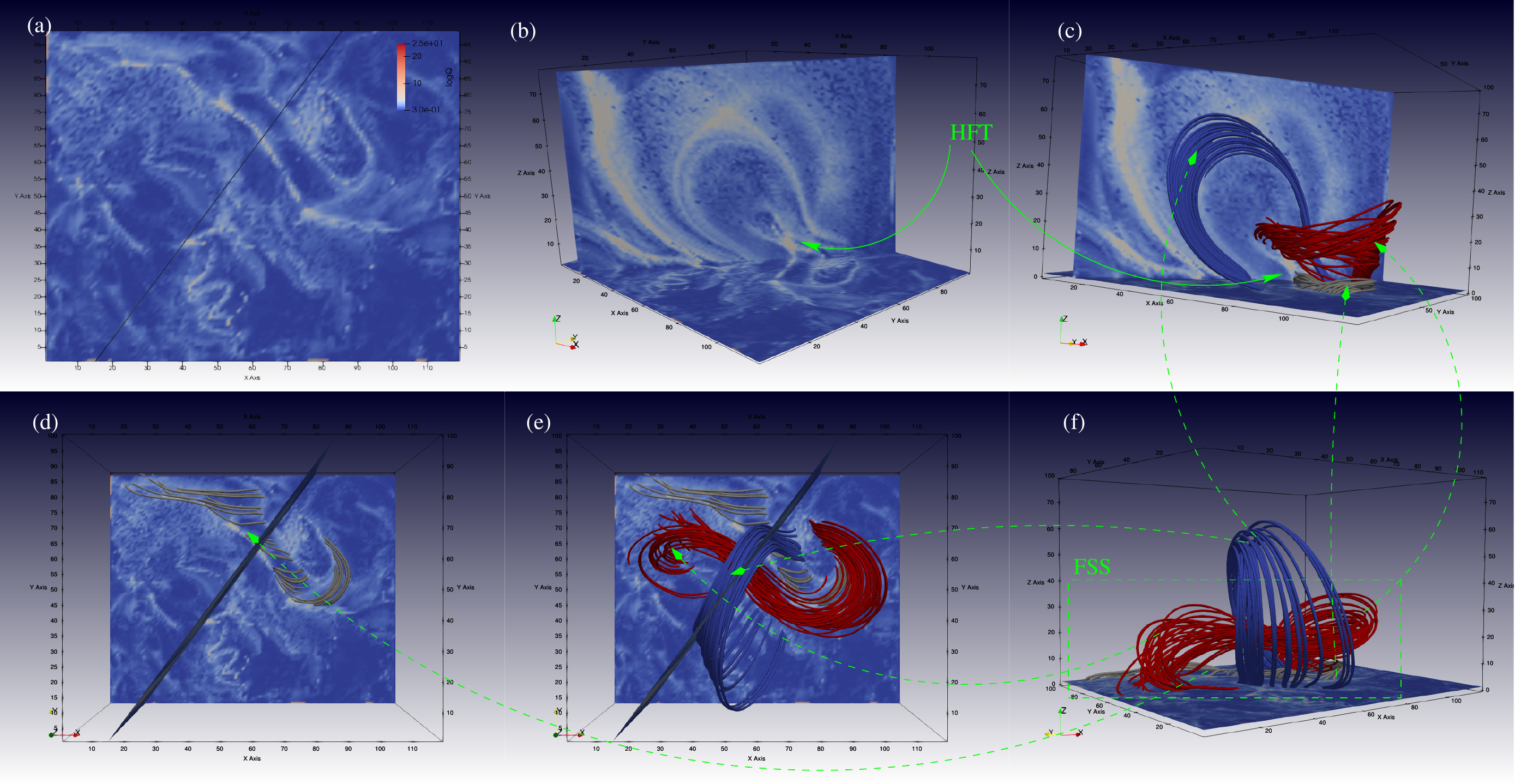}
	\caption{Same as Figure \ref{fig: qslf1} but for the SOL2013-04-11T event at 06:50 UT. Blue magnetic field lines that mimic the observed overlying flux bundles. } 
	
	\label{fig: qslf2} 
\end{figure*}
\begin{figure*}
	\centering \includegraphics[width=7in]{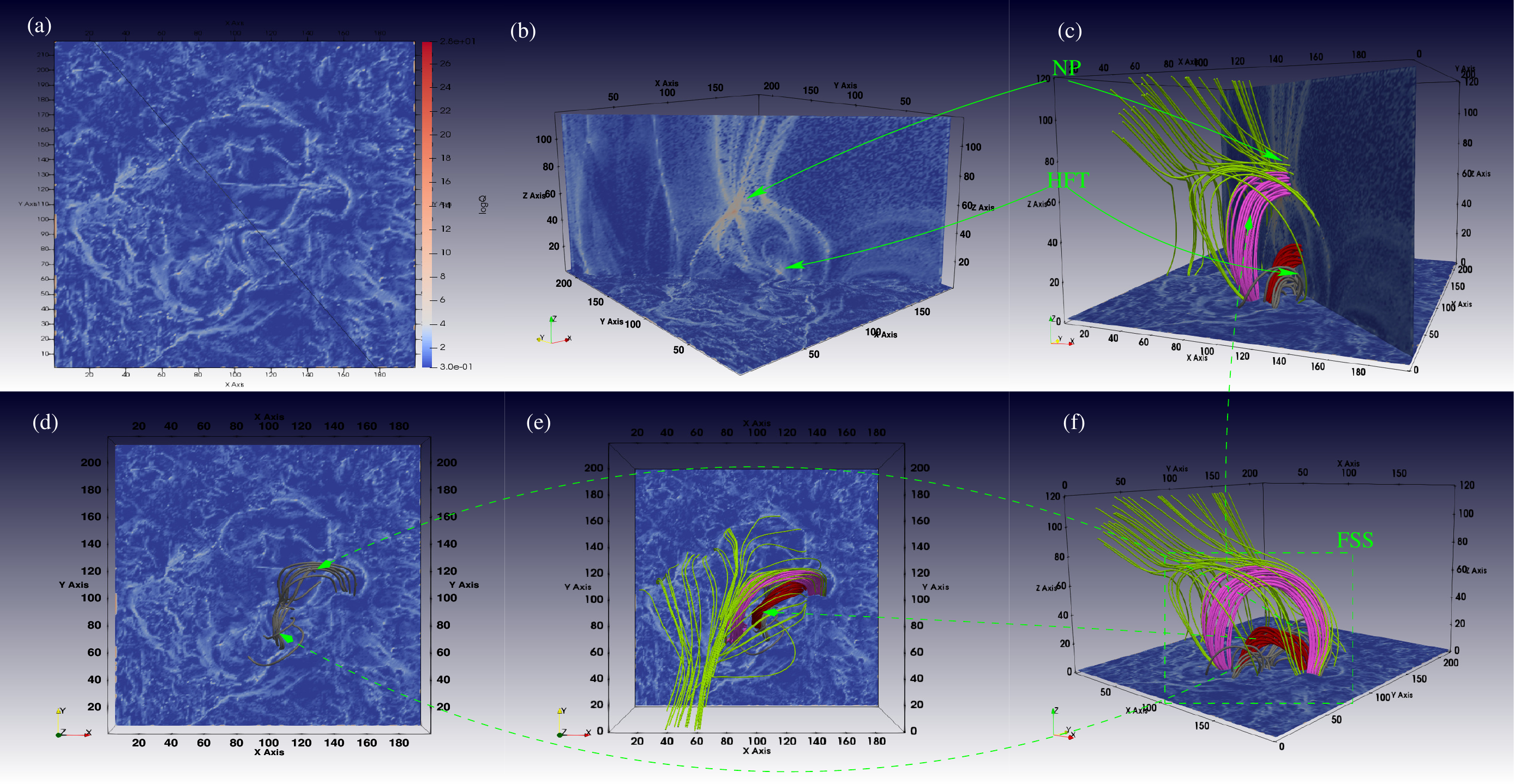}
	\caption{Same as Figure \ref{fig: qslf1} but for the SOL2014-04-18T event at 12:00 UT.  Note that the field lines in panel e appear to be different from those in panel of Figure \ref{fig: mstr3}, which is due to projection effects caused by different view angles of these two plots.} 
	
	\label{fig: qslf3} 
\end{figure*}
\begin{figure*}
	\centering \includegraphics[width=7in]{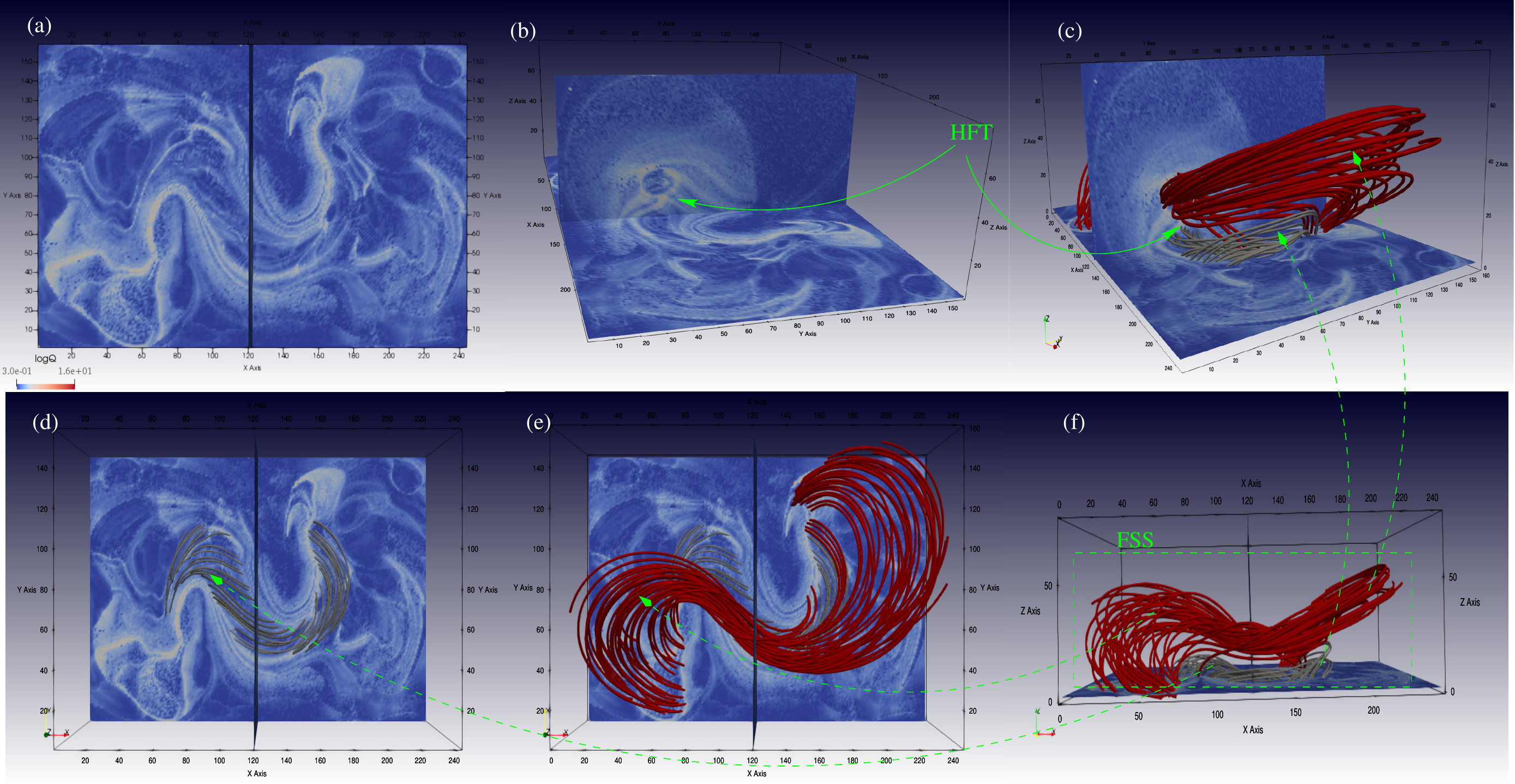}
	\caption{Similar to Figure \ref{fig: qslf1} but for the SOL2014-09-10T event at 17:20 UT.} 
	
	\label{fig: qslf4} 
\end{figure*}

Magnetic reconnection plays an important role in accumulation and release of magnetic energy in solar flares/CMEs. Confirming the locations of magnetic reconnection is crucial to determine the eruption mechanisms. Magnetic topology is defined as the unchangeable geometrical properties of the magnetic field and can be represented by the magnetic field line linkages. Quasi-sepatatrix layers (QSLs: \citealt{Priest_1995_jgr_100_23443-23464, Demoulin_1996_aap_308_643-655}) are defined as the places where the linkages are continuous but change drastically. In other words, QSLs are regions where magnetic gradient is large and thus the reconnection most likely takes place. They divide different magnetic domains and two QSLs converge at a location defined a hyperbolic flux tubes (HFT; \citealt{Titov_2002_JournalofGeophysicalResearch(SpacePhysics)_107_1164}). \cite{Titov_2002_JournalofGeophysicalResearch(SpacePhysics)_107_1164} proposed an invariant quantity called the squashing factor Q which is uniform along a magnetic field line to measure the mapping of magnetic field lines. Previous studies such as \cite{2015ApJ...810...96S, 2016ApJ...817...43S} and \cite{Janvier2016A&A...591A...141} have shown good correspondences between magnetic topology (e.g., QSL and HFT) and active region features (e.g., flare ribbons).

In order to understand the 3D topology of the source regions of the events we studied,  we adopt the code developed by \cite{Guo_2013_TheAstrophysicalJournal_779_157} and \cite{Yang_2015_TheAstrophysicalJournal_806_171} to calculate the squashing factor Q of the reconstructed magnetic fields. QSL maps at different views overlaid with selected field lines similar to those in Figures \ref{fig: mstr1}-\ref{fig: mstr4} are presented in Figures \ref{fig: qslf1}-\ref{fig: qslf4}, from which typical topology skelotons such as HFT, overlying null-point (NP), fan and spines are identified in one or more events. We find that two roughly parallel high Q (white) ribbons in the photospheric QSL maps (panel a) correspond to the footpoints of the HFTs displayed in panel b of Figures \ref{fig: qslf1}-\ref{fig: qslf4}. Magnetic field lines below the HFTs (gray) appear to be sheared arcades with footpoints located at the two parallel high Q ribbons as shown in panel d of Figures \ref{fig: qslf1}-\ref{fig: qslf4}, while magnetic field lines above the HFTs (red) are continuous and long S-shape that make up the MFRs with two footpoints located at the two ends of the high Q ribbons (panel e of Figures \ref{fig: qslf1}-\ref{fig: qslf4}e).

In particuar, the blue magnetic field lines that match well with the first erupted hot blobs in SOL2013-04-11T event  (marked by dashed yellow line in Figure \ref{fig: evo2}d ) are shown in Figure \ref{fig: qslf2}c, e and f. Overlying null points above the filament-sigmoid system are identified in the SOL2012-07-12T (Figure \ref{fig: qslf1}c) and SOL2014-04-18T (Figure \ref{fig: qslf3}b) events. Figure \ref{fig: qslf3}c, e and f show the fan-spine structure and the footpoints match well with the rectangular high Q ribbons which are consistent with the observed post-flare quasi circular flare ribbons  as shown in \citet{2015ApJ...812...50J}. The pink field lines in panels c and f of Figures \ref{fig: qslf3} above the flux rope near the overlying null point correspond to the observed hot channels.

\subsection{Formation of the Filament-Sigmoid Systems}\label{subsec:for}

\begin{figure*}
	\centering \includegraphics[width=7in]{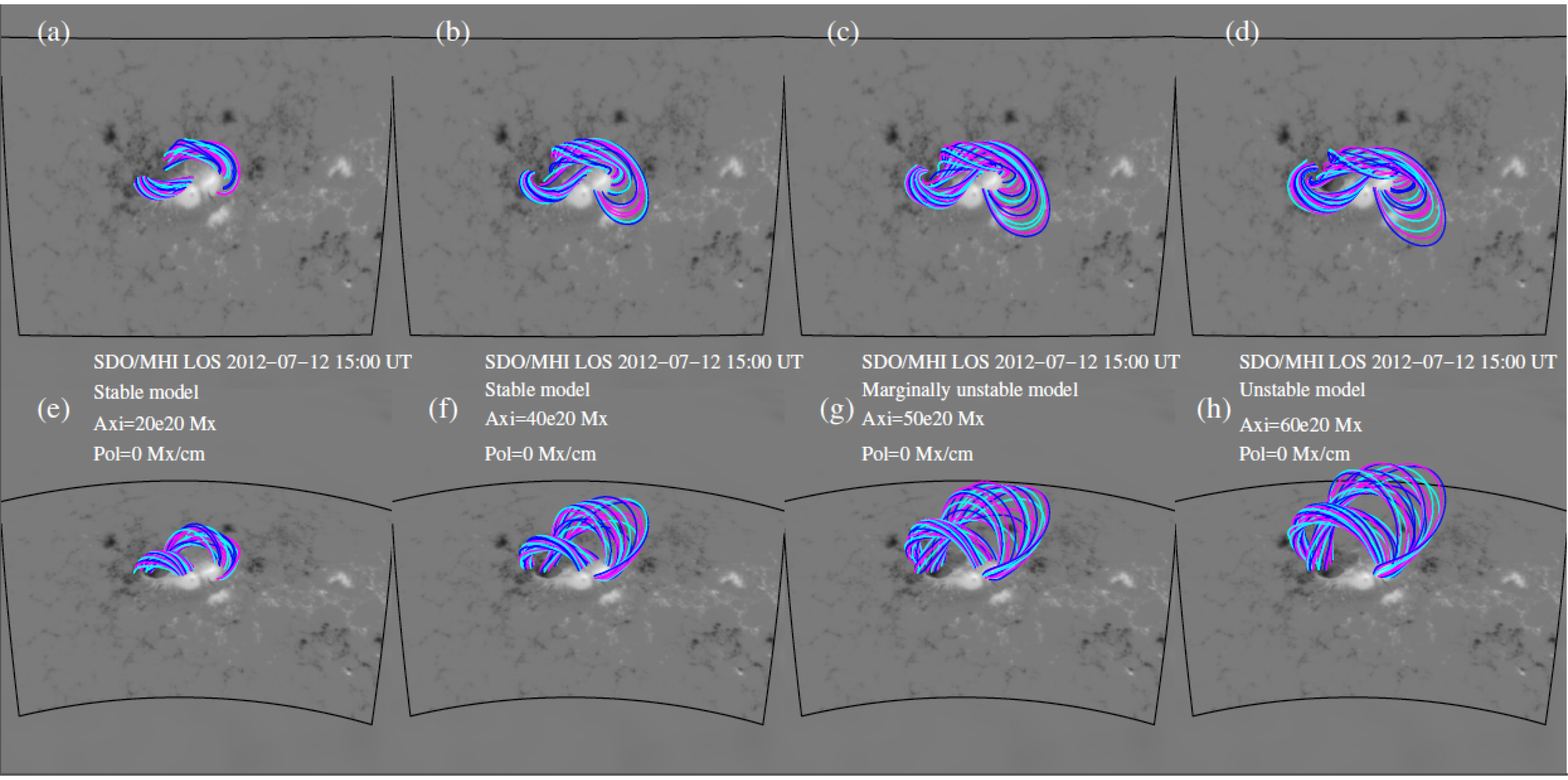}
	\caption{The background of these panels are the HMI Line-of-sight photospheric magnetograms (inside of the black rectangle) at 15:00 UT on 12 July 2012 and the HMI synoptic map (outside of the black rectangle). Multicolor curves represent magnetic field lines. Panels a-d show magnetic field lines in our models whose initial axial flux increases. The first two are stable models, the third is marginally unstable model and the last is unstable model, they are shown in the direction of observation. Panels e-h show the same magnetic field lines in another direction. Parameters of these models are shown in Tables \ref{tab:cha} and \ref{tab:par}.} 	
	\label{fig: mevo1} 
	\end{figure*}
\begin{figure*}
	\centering \includegraphics[width=7in]{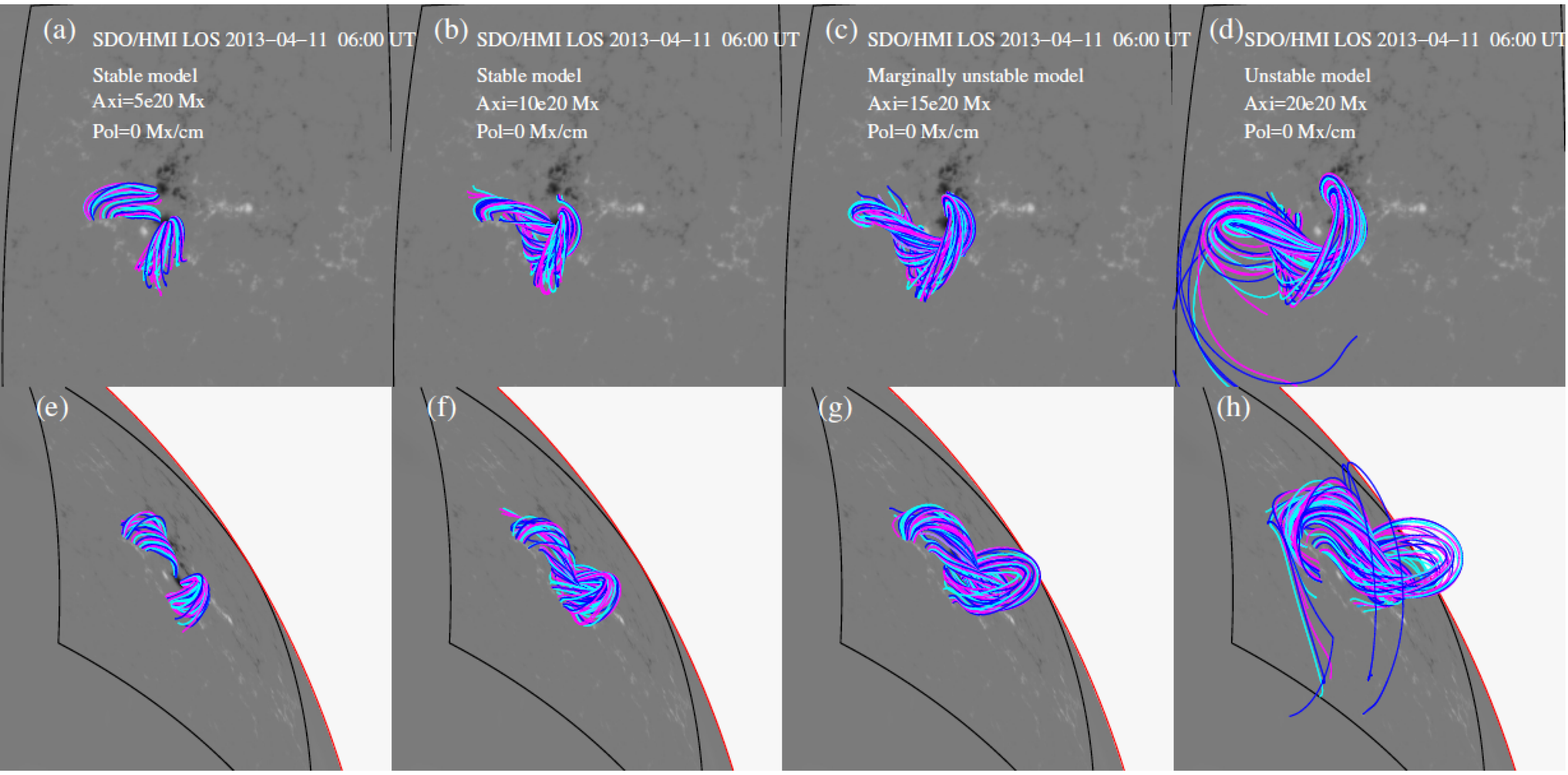}
	\caption{Same as Figure \ref{fig: mevo1} but for the SOL2013-04-11T event at 06:50 UT.} 
	
	\label{fig: mevo2} 
	\end{figure*}
\begin{figure*}
	\centering \includegraphics[width=7in]{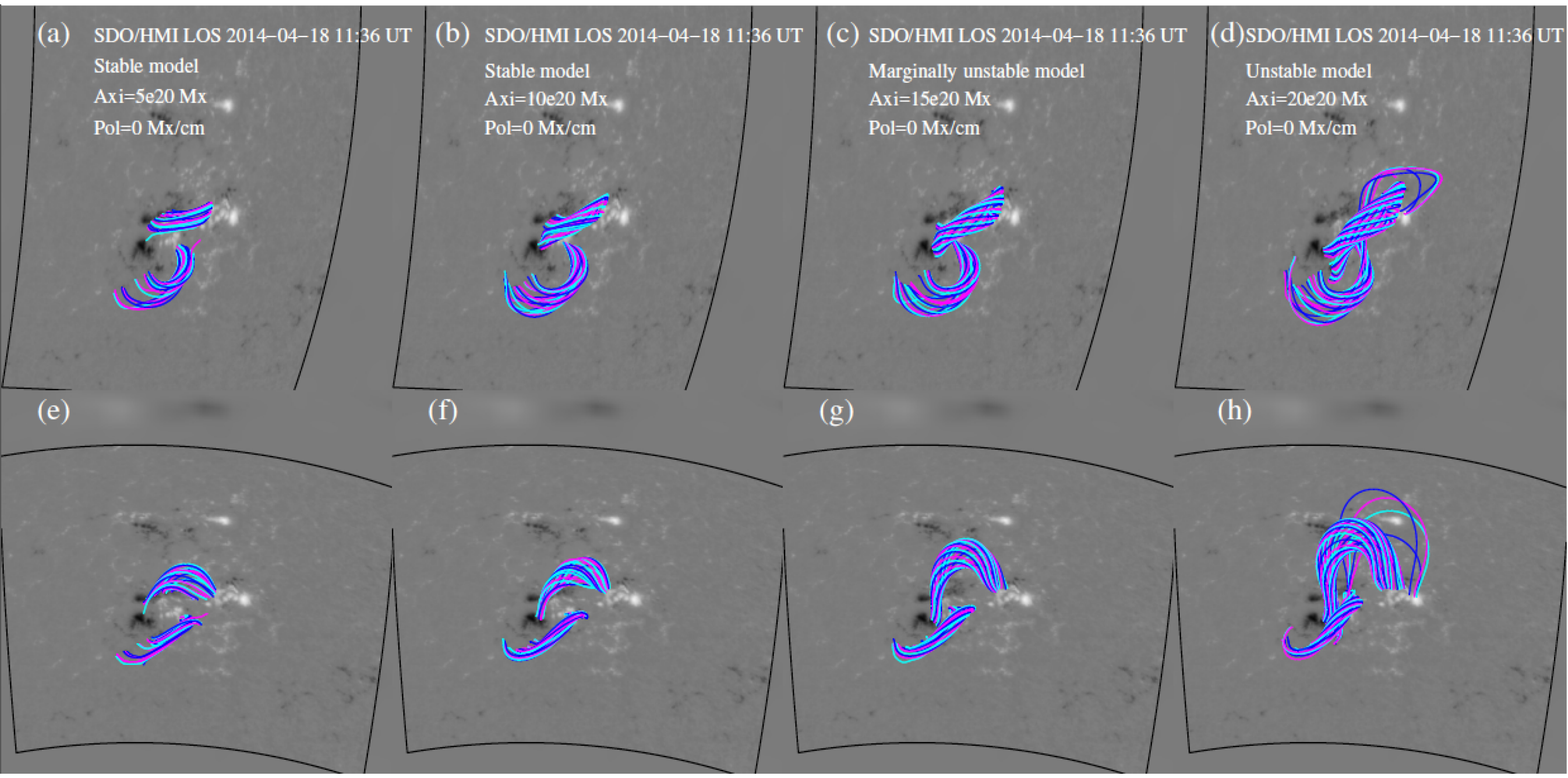}
	\caption{Same as Figure \ref{fig: mevo1} but for the SOL2014-04-18T event at 12:00 UT.} 
	
	\label{fig: mevo3} 
	\end{figure*}
\begin{figure*}
	\centering \includegraphics[width=7in]{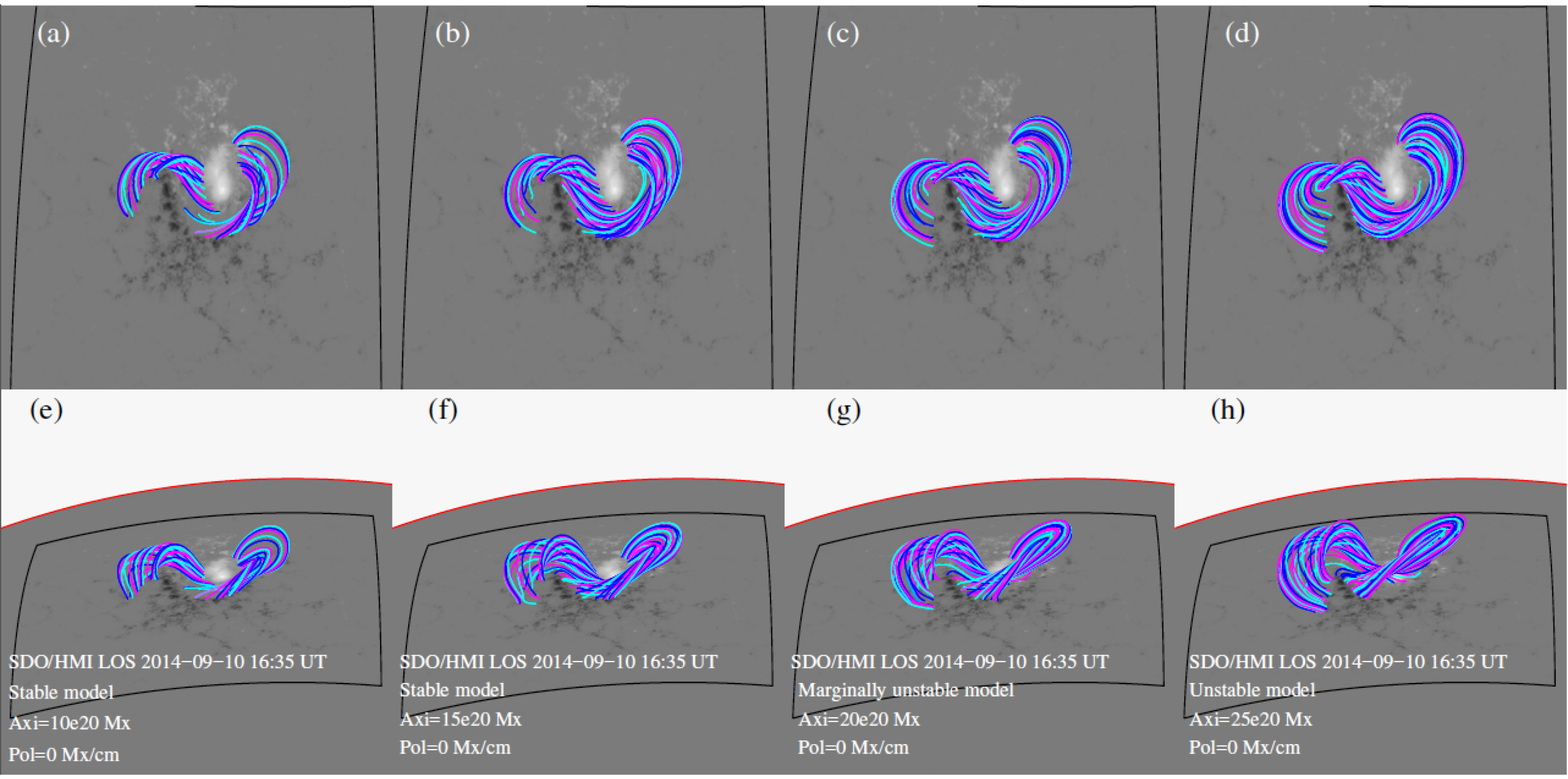}
	\caption{Same as Figure \ref{fig: mevo1} but for the SOL2014-09-10T event at 17:20 UT.} 
	
	\label{fig: mevo4} 
	\end{figure*}
	
In order to understand the formation of the filament-sigmoid system, we built a series of magnetic field models from stable to unstable by increasing the initial axial flux of the inserted flux rope.  Figures \ref{fig: mevo1}-\ref{fig: mevo4} show selected magnetic field lines of these models. The top row of Figures \ref{fig: mevo1}-\ref{fig: mevo4} shows field lines in the direction of observation and the bottom row displays the same field lines in another view. For each event, we present four models, the first two are stable models, the third one is a marginally unstable model, and the last one is unstable. Parameters of these models are listed in Table \ref{tab:par}. With the increase of axial flux, the models evolve from nearly potential to more non-potential state, during which double J-shape field lines merge into long continuous S-shape field lines. This is similar to the aformentioned observations in Section \ref{sec: Observations}  (Figures \ref{fig: evo1}-\ref{fig: evo4}). It suggests that the high-lying hot channel is more likely formed by increasing axial flux through magnetic reconnections such as tether-cutting reconnection. 

\begin{figure*}
	\centering \includegraphics[width=7in]{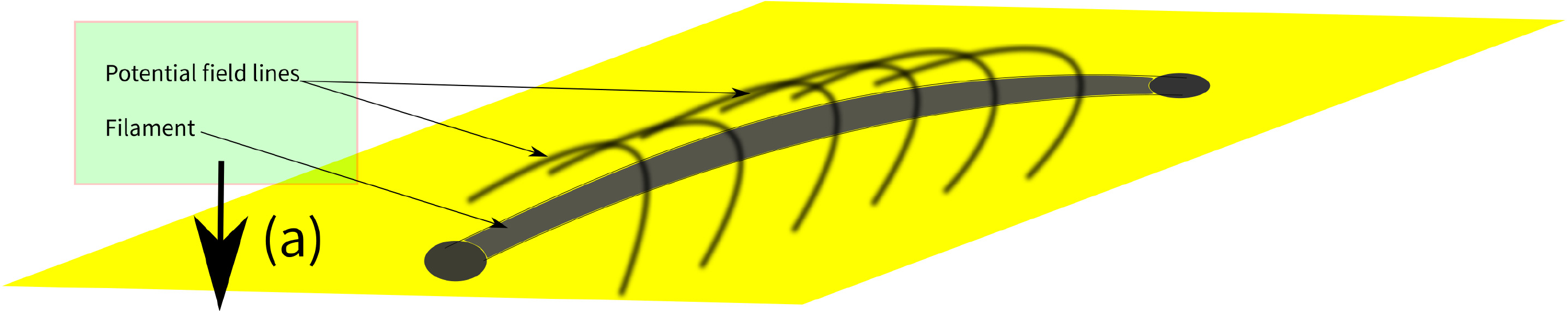}
	\centering \includegraphics[width=7in]{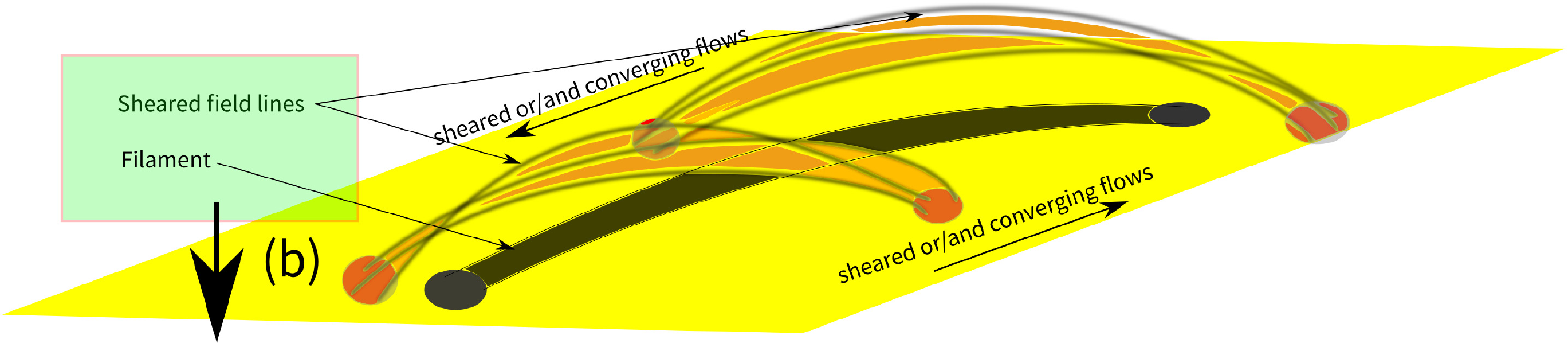}
	\centering \includegraphics[width=7in]{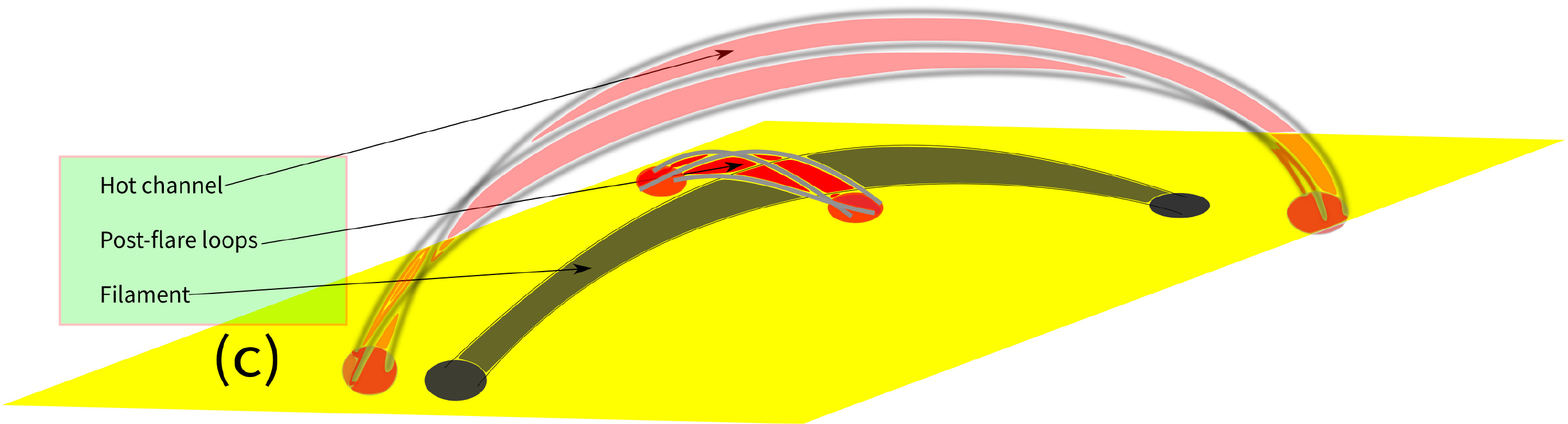}
	\caption{Schematic drawing of the formation of the filament-sigmoid system especially the hot channel. A filament and overlying potential field lines make up the initial configuration (panel a). Potential field lines become sheared arcades due to shear or/and converging flows as shown in panel b. Reconnection may already takes place at this stage. Panel c shows the formation of the high-lying hot channel and the brightenings surounding the filament channel through tether-cutting reconnection.} 	
	\label{fig: fil_cha} 
\end{figure*}

In this work, we consider the filament and hot channel as a filament-sigmoid system, the formation of which is illustrated in Figure \ref{fig: fil_cha}. At first, the region contains a low-lying filament with overlying potential fields (panel a). The overlying potential fields are gradually driven to be sheared due to processes such as shear or/and converging flows (panel b). Magnetic reconnection takes place in the sheared fieds above the low-lying filament, which leads to the formation of the high-lying hot channel and the low-lying post flare loops or/and brightenings surrounding the udnerlying filaments (panel c). The formation mechanism of the low-lying filament formed much earlier than the hot channel is out of the scope of this work.

\section{SUMMARY AND DISCUSSION} \label{sec:SUMMARY AND DISCUSSION}

\begin{figure*}
	\centering \includegraphics[width=7in]{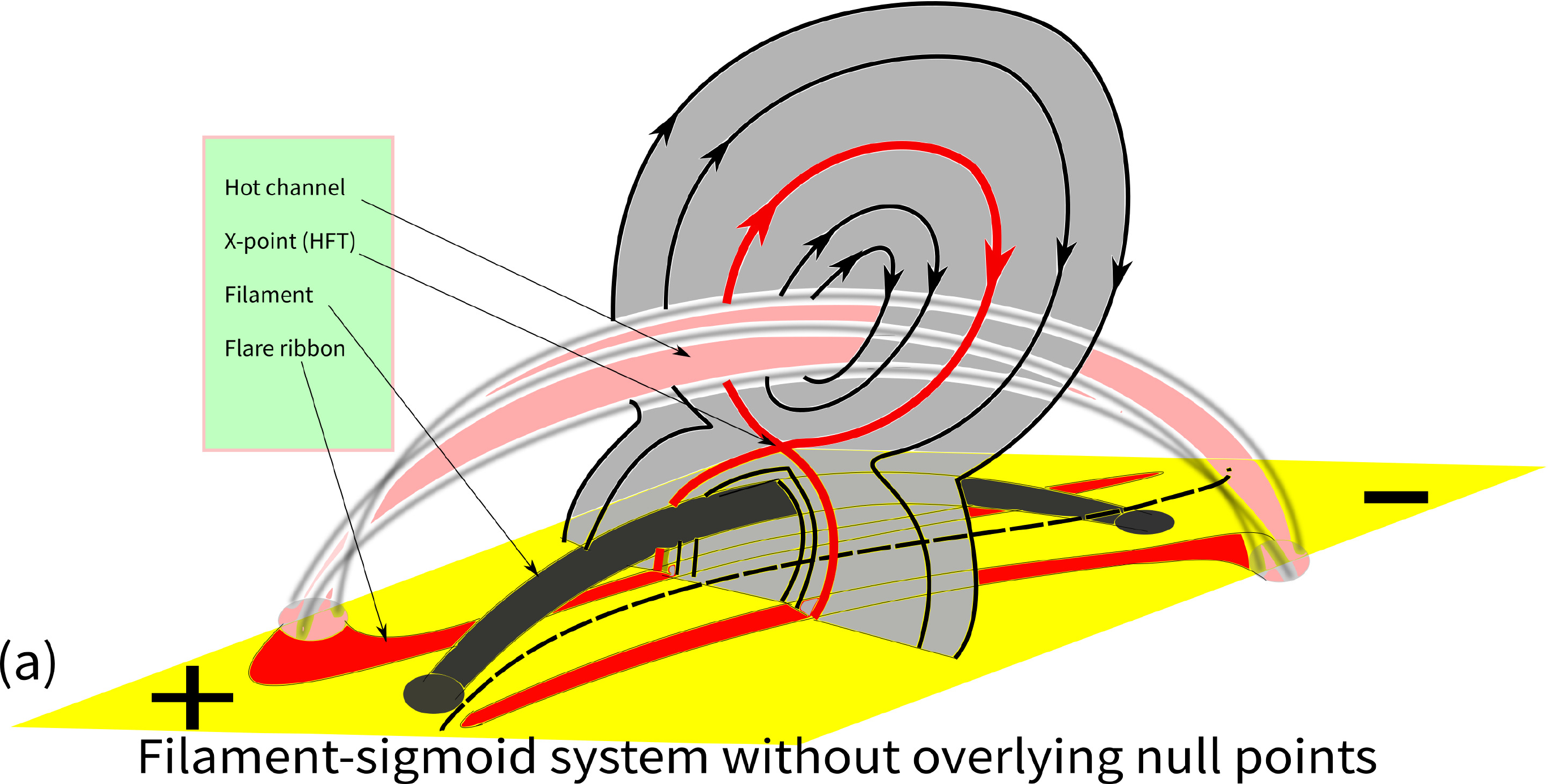}
	\centering \includegraphics[width=7in]{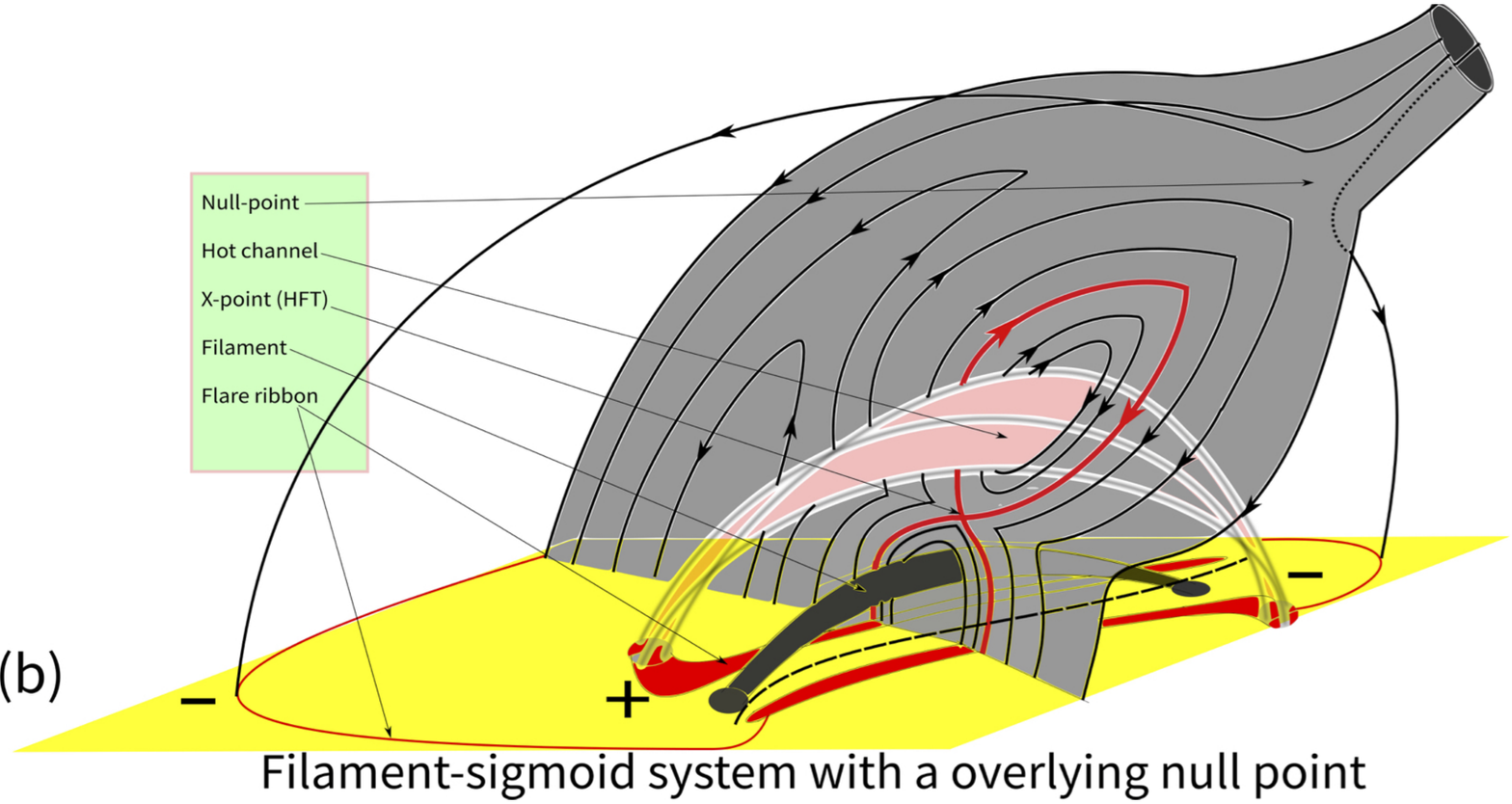}
	\caption{This cartoon shows the magnetic field structure of the filament-sigmoid system. Panels a shows a filament-sigmoid system without overlying null points and panel b presents a filament-sigmoid system with a overlying null point. We choose a cross-section (gray) which looks like a standing $\propto$ to show the HFTs. Filament and hot channel locate below and above the HFTs. Arrows represent directions of the magnetic field in this cross-section. The red arrowed curve is the separatrix of the MFR and the surounding magnetic field. PIL is displayed by the black dashed line. The double J shape channel (red) represents brightenings and/or flare ribbons in the chromosphere. The red curve in panel b reprensents the circular falre ribbons.} 	
	\label{fig: fil_cha5} 
\end{figure*}

In this paper, we study magnetic field configuration of four flare/CME events (SOL2012-07-12T, SOL2013-04-11T, SOL2014-04-18T, and SOL2014-09-10T), which occurred in four different active regions with high-lying sigmoidal hot channels and low-lying filaments that make up the filament-sigmoid systems. Hot channels in the first and the third events display S-shape, while inverse S-shape hot channels appear in the other two events. The eruption of all of these hot channels are followed by CMEs and flares. Comprehensive observational study of the hot channels in these four events has been carried out by \cite{Cheng2016} who are unable to obtain magnetic field lines that fit the observed hot channels using normal NLFFF extrapolations. In order to understand the magnetic structure of hot channels, we construct a series of magnetic field models for these active regions using the flux rope insertion method through adjusting axial flux and/or poloidal flux of inserted flux rope and carry out magnetic field topological analysis. Our main results and discussions are summarized as follows.

 AIA observations show that two groups of sheared arcades gradually evolve into a sigmoidal hot channel during tens of (or several) hours before the flare. Once the hot channel forms, it is usually stable for a while and then erupts suddenly with its footpoints moving towards further locations. During the formation of the hot channel, the preexisting underlying filaments remains barely changed. Our modeling shows that the same set of field lines in different models with increasing axial flux of inserted flux rope can mimic the observed formation process of hot channels, i.e., two groups of J-shape field lines merge into one group of long sigmoidal field lines. This is consistent with the previous observational findings for these four events that the continuous sigmoidal hot channel is built up from two groups of sheared arcades near the main polarity inversion line  before the eruption \citep{2014ApJ...789...93C, 2015ApJ...804...82C, 2015ApJ...812...50J, 2017ApJ...834...42J}, which is also similar to the simulation on the formation process of sigmoidal flux ropes for other events in the literature \citep{2009ApJ...703.1766S, 2012ApJ...759..105S, 2014ApJ...780...55J}.  In our work, we consider the filament and sigmoid as a system, and the hot channel is formed  by increasing of axial flux due to magnetic reconnections such as tether-cutting reconnection \citep{1984SoPh...94..341S, 2001ApJ...552..833M} occurred highly above the low-lying filament.

Observational study by \citet{Cheng2016}  suggests that the hot channel has ascended to a high altitude due to the observed significant deviation between the axis of the hot channel and the PIL (or the associated filament). The hot channels contain high temperature plasma since they appear in the AIA high temperature passbands. \cite{2012NatCo...3E.747Z} and \cite{2013ApJ...763...43C} speculate that the coherent channel-like hot structures are MFRs which exist prior to the eruption. Through magnetic modeling of the four flare/CME events, we find that the best-fit models composed of an MFR with HFT where magnetic reconnection most likely occurs can represent the filament-sigmoid systems prior to the flare onset. The height of the X-point/HFT is in the range of 4.9-16.8 Mm, and the height of the flux rope axis is located in a range between 19.8 Mm and 46 Mm above the photosphere. The sigmoidal hot channels  correspond to magnetic field lines located above the HFT; while those below the HFT surround the low-lying filament. In particular, the continuous and long field lines representing the flux rope located above the HFT are corresponding to the hot channels in three events. While the flux bundle that mimics the observed hot channel in SOL2014-04-18 are located above the flux rope.  It is worth noting that the NLFFF model built for the 2014-09-10T event using Grad-Rubin method  \citep{2016ApJ...823...62Z}  and the model built for SOL2012-07-12 event by \citep{2014ApJ...789...93C} also contain an MFR with HFT.  The heights of the HFT and flux rope axis in the first and second cases are slightly or much lower than those in our models, respectively. 
 
As time goes on, the high-lying hot channel rises faster then separates from the low-lying filament. In the end, the high-lying hot channels erupt and lead to CMEs and flares, while the low-lying filaments remain in full or in part. This behavior is similar to the so called partial eruptions \citep{2006ApJ...637L..65G, 2007SoPh..245..287G}. The only difference may be that the MHD simulations by \citep{2006ApJ...637L..65G} suggest that the partial eruption is caused by the splitting of one flux rope. Figure 1 of \cite{2006ApJ...637L..65G} shows that the surviving parts of the MFR are the set of dipped field lines and flux rope field lines that graze the bald-patch (BP) as the bald-patch separatrix surface (BPSS) below the reconnection cusp, and the escaping parts are the newly formed field lines of reconnection above the reconnection regions. However, in the filament-sigmoid system we think that the high-lying hot channel might be formed by the merging of two sheared arcades overlying the filament rather than splitting of one flux rope.  Moreover, the magnetic structures derived from our current models suggest that  the lowing filament is more likely to be supported by sheared arcades rather than a flux rope. However, we are not able to exclude the possibility of the existence of a low-lying flux rope as suggested by
\citet{2014ApJ...789...93C}, in which Figure 9 shows many BPs along the PIL below the filament for the SOL-2012-07-12 event. 

For the SOL2012-07-12 event, some authors like \cite{2014ApJ...789...93C} think that the AR may contain a double-decker flux rope including a high-lying MFR and a low-lying MFR. \citet{Liu_2012_TheAstrophysicalJournal_756_59} performs an observational study of a double-decker filament and suggests two types of force-free magnetic configurations that are compatible with the data, a double flux rope equilibrium and a single flux rope situated above a loop arcade. The follow-up work \citep{Kliem_2014_TheAstrophysicalJournal_792_107} carried out MHD simulations and find that a double-decker flux rope can lead to eruptions under conditions typical of solar active regions. However, no magnetic field reconstructions based on observational magnetic fields can reproduce this double-decker flux rope structure so far, neither do our current modeling. In our model, magnetic field lines below the HFT appear to be consistent the observed  low-lying filament and the high-lying hot channel is represented by field lines above the HFTs. This suggests that the observed double-decker structure may be separated by the HFTs, which is a representation of filament-sigmoid system similar to Figure 12b of \cite{Liu_2012_TheAstrophysicalJournal_756_59}. The heights of the X-point and the flux rope axis are 16.8 Mm and 44.8 Mm, which are much higher than those in the other three events. This might explain why this event is considered as a double-decker MFR structure.

 The pre-flare magnetic configurations in these four events are divided into two groups: filament-sigmoid system without overlying null points and filament-sigmoid system with an overlying null point. A sketch of the basic magnetic structure of these two groups of events is presented in Figure \ref{fig: fil_cha5}. The height of the overlying null point is  about 84.5 Mm and 55.3 Mm for SOL2014-04-18T event and SOL2012-07-12T event, respectively. Null points usually refer to a location where $B=0$ in the fan-spine configuration (\citealt{1990ApJ...350..672L, 2011ApJ...728..103L}). Magnetic reconnection may take place at the null point, which opens the overlying magnetic field and leads to eruptions such as CMEs and jets.   \cite{1999ApJ...510..485A} and \cite{2000ApJ...540.1126A} proposed the ``magnetic breakout" model whose central idea is that reconnection at the overlying null point removes background magnetic fields.  For SOL2014-04-18T event, \cite{2015ApJ...812...50J}  provide several observational evidences such as the existence of a closed dome-like structure and a large scale circular ribbon to support the existence of an overlying null point which is not identified in their NLFFF models.  However, an overlying null point with fan-dome topology is identified in our magnetic field model. The best-fit models for the two events with overlying null points (SOL2012-07-12T and SOL2014-04-18T ) are marginally unstable models.  While the best-fit model for SOL2014-09-10T event without overlying null point is an unstable model. A marginally unstable model is also identified as the best-fit model for the SOL2013-04-11T event, and the first erupted overlying flux bundle might play a role in the initiation of this event as suggested by \citet{2014ApJ...797...80V} .  ``Marginally unstable state" refers to a state at the boundary between stable and unstable states in parameter space. When the magnetic structure reaches marginally unstable it is easy to erupt, and if the model is unstable, the ideal MHD instabilities are most likely to occur. In our previous work\citep{2009ApJ...691..105S,2009ApJ...704..341S,2011ApJ...734...53S}, we often find that the best-fit preflare models are generally stable, unlike which the modeling in the current work suggests that prior to eruption each region is already marginally unstable, and a small disturbance such as tether-cutting/breakout reconnection  can lead to the explosive eruption. The existence of an overlying null point might play a role in the instability of the region. Further analysis are required to obtain conclusive statements.

We acknowledge Drs. Bernhard Kliem and Rui Liu for valuable suggestions and comments to improve the study. We are grateful to Dr. Yang Guo for providing the code to perform the topological study and helpful discussions. We thank the team of SDO/AIA and SDO/HMI for providing the valuable data. The HMI and AIA data are downloaded via the Virtual Solar Observatory and the Joint Science Operations Center. This work is supported by NSFC 11473071, 11333009, 11790302 (11790300) and U1731241. Yingna Su is also supported by the One Hundred Talent Program of the Chinese Academy of Sciences (CAS) and the Natural Science Foundation of Jiangsu Province (Grant No. BK20141043). This work is also supported by the Strategic Priority Research Program on Space Science, CAS, Grant No. XDA15052200 and XDA15320301. Xin Cheng is supported by NSFC through grants 11722325, 11733003, 11790303. Xin Cheng is also supported by Jiangsu NSF through grants BK20170011 and “Dengfeng B” program of Nanjing University.

\bibliographystyle{aasjournal}
\bibliography{thefour}

\end{document}